# The Effect of Age on the Grouping of Open Clusters: The Primordial Group Hypothesis

Juan Casado

Facultad de Ciencias, Universidad Autónoma de Barcelona, 08193, Bellaterra, Catalonia, Spain

Email: juan.casado@uab.cat

**Abstract:** The Primordial Group hypothesis states that only sufficiently young open clusters (OCs) can be multiple, and old OCs are essentially isolated. We test such postulate through four different studies using a manual search of Gaia EDR3 and extensive literature. First, we revisit the work of de La Fuente Marcos & de La Fuente Marcos (2009), which states that only ca. 40% of the OCs pairs are of primordial origin. However, no plausible binary system among their proposed OC pairs having at least one member older than 0.1 Gyr has been found. Second, we research the OCs < 0.01 Gyr old in Tarricq et al. (2021) and obtain that ca. 71% of them remain in their primordial groups. Third, a similar study of the oldest OCs (age > 4 Gyr) shows that they are essentially alone. Forth, the well-known case of the double cluster in Perseus and some other literature binary systems are shown to accommodate the title hypothesis, too. A simplified bimodal model allows retrieving the overall fraction of related OCs (around 12-16 %) from our results, assuming that young clusters remain associated ~0.04 Gyr. The obtained results further support that OCs are born in groups (Casado 2021).

**Keywords:** Open Cluster Pairs; Open Cluster Groups; Open Cluster Formation; Gaia; Manual Search; Primordial Group Hypothesis

## 1. Introduction

Open clusters (OCs) are born from the gravitational collapse of gas and dust in giant molecular clouds. There is observational evidence that some of them are born in groups (Bica et al. 2003; Camargo et al., 2016). Galactic OCs are sometimes found in pairs, and the number of these optical pairs is significantly higher than would be expected if clusters were randomly distributed (e.g., Rozhavskii et al. 1976). Studies of grouping among OCs provide keys to understanding star formation in the Galactic disk and the subsequent dynamical evolution of OCs.

Until recently, h and $\chi$ Persei was the only accepted physical double cluster in our Galaxy (Vázquez et al. 2010), even though some literature on other OC pairs already existed (e.g., Subramaniam et al. 1995; Piskunov et al. 2006; de La Fuente Marcos & de La Fuente Marcos 2009). Conversely, roughly 10% of the known OCs in the Large Magellanic Cloud (LMC) seem to belong to pairs (Dieball et al. 2002). Back to the Galactic disk, the first estimations of the fraction of these paired clusters by statistical comparison of entirely random distribution of OCs or pairs of OCs reached a level of 20% (Rozhavskii et al. 1976). Bica et al. (2003) estimated that a fraction of 25% of the embedded clusters (EC) is formed in pair or triplet systems, although groups with up to 9 OCs were also identified (see Casado 2021). Subramaniam et al. (1995) estimated that about 8% of the OCs in the Galaxy appear to be members of binary systems. De La Fuente Marcos & de La Fuente Marcos (2009) argued that the real fraction was similar to that in the LMC. Out of this population, nearly 40% of them were classified as primordial binary open clusters. However, only about 17% appear to survive for more than 25 Myr (de La Fuente Marcos

& de La Fuente Marcos 2010). On the other hand, Soubiran et al. (2018) used *Gaia* data sets and analyzed the 6D space-phase volume. They recognized only five likely binary clusters and a group of 5 OCs differing by less than 100 pc in their Galactic position and 5 km/s in velocity from their high-quality sample of 406 OCs. Nevertheless, in a corrected version of the same work, Soubiran et al. (2019) listed 21 candidate pairs. Anyhow, the exact fraction of related OCs in the Galaxy remains unknown at present. However, new precision *Gaia* data will hopefully ascertain them soon, at least for the solar neighborhood.

The second *Gaia* Data Release (*Gaia* DR2) provided precise astrometric data (position, parallax (plx), and proper motions (PMs)) and (1+2)-band photometry for about 1.3 billion stars (Gaia Collaboration, 2018), so starting a new era in precision studies of Galactic OCs (among other subjects). The recent third release of *Gaia* early data results (*Gaia* EDR3; Gaia Collaboration, 2020) improves, even more, the accuracy of the measurements for around 1.5 billion sources.

Today, virtually all new OCs are found through some unsupervised algorithm that detects overdensities in high-dimensional space from the plethora of data in large stellar databases and information provided by space missions such as *Gaia* (e.g., Cantat-Gaudin et al. 2018; Castro-Ginard et al. 2020). The next step of validation and characterization is performed using automatic machine learning techniques. For instance, the approach applied by Castro-Ginard et al. (2020) has detected hundreds of new OCs in the *Gaia* database. However, this method does not recover a fraction of the OCs from the literature, partly due to the non-existence of several of these OCs (Cantat-Gaudin & Anders 2020). This fact suggests that their approach may also be unable to detect a fraction of undiscovered OCs (Hunt & Reffert 2020). However, two recent studies have identified dozens of previously unknown OCs via manual mining of the *Gaia* dataset (Casado 2020, 2021). These surveys are less productive in the quantity of new OCs and cannot ensure completeness, either. However, they are more detailed since the manual approach allows going beyond the purely formal search of OCs: Each OC candidate is examined individually, based on extensive available data.

In one of those studies (Casado 2021), a comprehensive list of 22 double or multiple OCs comprising 80 possible member clusters between the galactic longitudes of 240$^{o}$ and 270$^{o}$ has been examined with the help of *Gaia* EDR3 and the existing literature. We discovered that almost all the 52 most likely grouping members are OCs younger than 0.1 Gyr. We did not find any likely groups containing older OCs. These results suggest that most groups, if not all, are of primordial origin and are not stable for a long time, in line with similar conclusions obtained from the study of the Magellanic Clouds (Hatzidimitriou & Bhatia 1990; Dieball et al. 2002). Our results also suggest a low probability, if any, of pairs formed by tidal capture or resonant trapping, which would be due to the small likelihood of close encounters of OCs, and the even lower probability of tidal capture without disruption of at least one of the clusters. Estimations of the fraction of OCs that form part of groups (9.4 to 15%) support the hypothesis that the Galaxy and the Large Magellanic Cloud are similar in this respect, too (de La Fuente Marcos & de La Fuente Marcos 2009). One of our conclusions was that OCs are generally born in groups, i.e., in clusters of clusters. The stellar formation process is being depicted as more complex than previously thought. See, for instance, the complex structures derived in Vela-Puppis (Cantat-Gaudin et al. 2019) and the case of Group C in this paper. Other studies suggest a similar scenario (de La Fuente Marcos & de La Fuente Marcos 2009b; Grasha et al. 2017; Kounkel & Covey 2019). Groups of young OCs are the likely result of such a hierarchical and turbulent star formation process.

In the present work, we propose and test the dubbed *Primordial group* hypothesis: i.e., only young enough OCs can be multiple, and old OCs are essentially single, since the gravitational interaction between OCs in primordial groups is indeed weak, and the probability of gravitational capture of OCs originated in different molecular clouds is very low (Dieball et al. 2002). We do it by four diverse studies. In section 3, we review some cluster pair candidates proposed by de La Fuente Marcos & de La Fuente Marcos (2009). In section 4, we look for companions of the clusters younger than 0.01 Gyr, and in section 5, the same method is applied to OCs older than 4 Gyr. In section 6, we revisit the case of the double cluster in Perseus and a few other literature binary cluster candidates, and section 7 summarizes some concluding remarks.

In the present study, pair (or group) of OCs refers to any candidate group of interacting OCs, whether gravitationally bound or not, while the term "binary cluster" is limited to gravitationally bound OC pairs.

## 2. Methodology

The methods applied to find and select pairs of OCs have been detailed previously (Casado, 2021). However, we recapitulate here the general methodology.

We start with a candidate member of a hypothetical pair (or group) of OCs. For each of these candidates, we look for close correlations between coordinates, PMs, and parallax for all OCs within the studied area (at least 100 pc around each studied OC). For example, if two OCs are close enough (i.e., at a projected distance of fewer than ten times the smaller of their radii and less than 100 pc), the rest of their astrometric data are compared. If there is any overlap of the data, considering uncertainty intervals of 3σ, both OCs are included in Table 2. The table is refined using the most accurate and recent parameters on individual OCs from reported studies using Gaia DR2 and Gaia EDR3, where possible. When the existing data are questionable, incomplete, or inconsistent between different authors, the OCs are manually re-examined using Gaia EDR3 to obtain the corresponding parameters. The Gaia EDR3 astrometric solution is accompanied by new quality indicators, such as the renormalised unit weight error (RUWE). RUWE allows sources with inaccurate data to be discarded (Gaia Collaboration 2020). We have routinely discarded sources that have RUWE > 1.4. Unless otherwise stated, we have also discarded sources of *Gmag* > 18 to limit parallax and PM errors, which increase exponentially beyond this magnitude threshold. The member stars of each new or re-examined OC are obtained through an iterative method that has been detailed previously (Casado 2020). In summary, this method refines the approximate allowed ranges in position, PM and plx, initially obtained by eye, by examining the resulting CMD, which must include a maximum number of likely member stars but a negligible number of outliers (stars out of the OC′ evolutionary sequences on its CMD). The error ranges in Tables 1 and 2 are not the standard uncertainties, but the absolute (maximum) errors that encompass all the member stars of each OC.

Following the criteria of previous studies, the obtained groups were refined by discarding OCs that are more than 100 pc away from any other member (Conrad et al. 2017; Soubiran et al. 2018; Liu & Pang 2019; Casado 2021), assuming that all members are at the average distance (*d*) of the group. This cutoff is an order of magnitude and an unrestrictive maximum, as other studies have used more restrictive limits (e.g., de La Fuente Marcos & de La Fuente Marcos (2009) used 30 pc). Groups with differences in radial velocities (RV) >10 km/s were also discarded (Conrad et al. 2017; Casado 2021). Other requirement for refining the groups is that ΔPM/plx (or ΔPM *d*) is < 2 $yr^{-1}$ between each pair of group members, using the units in Tables 1 and 2. The latter condition implies that the differences in tangential velocities are also less than 10 km/s. Some limiting cases for each group are discussed in the following sections.

A straightforward way to search for OC candidates linked to each studied OC is to plot a graph of the Gaia sources satisfying the examined OC constraints for the studied field. In this way, we can obtain plots similar to Fig. 1, showing (or not) the associated OCs. These charts are free of most of the noise from the unrelated field stars.

The data in the pre-Gaia literature are significantly less precise than those in Gaia, especially when talking about PMs (generally excluded from the analysis) and ages (Paunzen & Netopil 2006). Nevertheless, most of the reported data on *d*, RV, and even age of well-studied OCs have some value. Therefore, they have been used in the indi-vidual discussion of candidate pairs to compare and confirm *Gaia* data or when no *Gaia* studies have been found, as is the case for some of the reexamined OCs.

## 3. Analysis of candidate OC pairs from de La Fuente Marcos & de La Fuente Marcos (2009)

De La Fuente Marcos & de La Fuente Marcos (2009) (DFM hereafter) proposed numerous pairs of Galactic OCs and stated that only ~ 40% of cluster pairs are probably primordial. On the other hand, the statistics in Casado (2021) suggest that the vast majority of candidate pairs and groups found are probably primordial. To test the *Primordial group* hypothesis, in this section, we reexamine the candidate pairs of DFM having at least one member older than 0.1 Gyr. If we find some binary systems that have at least one member > 0.1 Gyr, as suggested by DFM, the proposed hypothesis would be falsified.

The candidate pairs in DFM were only selected from their position and distance, which must be compatible with a projected distance between them of less than 30 pc. We perform a deeper study using data from extensive literature and *Gaia* EDR3. The main results are summarized in Table 1.

DFM considered OC data from two catalogs: the WEBDA *Open Cluster Database* (Mermilliod & Paunzen 2003) and the *New Catalogue of Optically Visible Open Clusters and Candidates* (NCOVOCC; Dias et al. 2002). In the following subsections, we discuss our findings on each candidate pair comprising any member older than 0.1 Gyr.

### *3.1. WEBDA Catalog*

Pair #1 - ASCC100/ASCC101

DFM classified Pair #1 as a hyperbolic encounter, i.e., not a true binary system. Moreover, according to DFM, Pair #1 could be controversial: the member objects may not be real OCs. Despite ASCC100 was not found by the algorithm of Cantat-Gaudin et al. (2018), our manual study of ASCC 100 and the existing literature confirm the existence of both OCs, although not their binary nature. Positions, parallaxes, and RVs are compatible, but PMs are disparate (Table 1). The difference in the mean parallaxes can be ascribed to a global offset of *Gaia* DR2 parallaxes, which are 0.029 mas too small on the whole (Lindegren et al. 2018). Reported RVs for ASCC 100 range from -22.9 km/s (Kharchenko et al. 2005) to -25.9 km/s (Dias et al. 2002), while for ASCC 101, RVs span from -15 km/s (Tarricq et al. 2021) to -32 km/s (Conrad et al. 2017). However, these similar RVs seem to merely reflect the general motion of the stars in that particular region of the Galaxy. ASCC 100 is a relatively young OC, with all reported ages in the narrow interval from 0.089 Gyr (Dib et al. 2018) to 0.102 Gyr (Vande Putte et al. 2010). However, ASCC 101 is a mature OC, which reported age spans from 0.33 Gyr (Vande Putte et al. 2010) to 0.49 Gyr (Cantat-Gaudin et al. 2020). Although not determining, the different ages also suggest a chance encounter in the space of these otherwise unrelated OCs. The ensemble of results is consistent with the hyperbolic character of this pair. Note that, in hyperbolic encounters, the 3D positions should be close, but the 3D kinematics and ages should not.

Incidentally, Soubiran et al. (2018) found ASCC101 as a possible binary with NGC 7058, but ASCC 101 and NGC 7058 are 185 pc apart according to these authors, making this pair highly unlikely. Moreover, the corrected version of this paper (Soubiran et al. 2019) does not include this pair in the final list of candidates.

Pair #5 - ASCC 90/NGC6405

The *Gaia* DR2 mean parallaxes of ASCC 90 and NGC 6405 differ markedly, about 27%, and the photometric distances are ~22% apart (Table 1). From the celestial coordinates and the distances (the more conservative approach), we infer that both OCs are more than 100 pc apart, and the proximity condition would be not fulfilled. PMs in declination are only marginally compatible. The RV of ASCC 90 is in the interval 6.7 to 10.7 km/s, from seven member stars measured by *Gaia* DR2 (Tarricq et al. 2021). However, NGC 6405 has at least six consistent RVs from -7.0 km/s (Vande Putte et al. 2010) to -9.8 km/s (Loktin & Popova 2017). Thus, the difference between mean RVs is ca. 17 km/s, significantly higher than the accepted threshold of 10 km/s. Unsurprisingly, their galactic orbital parameters do not fit. For instance, the eccentricity is more than one order of magnitude higher for ASCC 90 than for NGC 6405 (Tarricq et al. 2021). All in all, the physical link of this candidate pair of OCs appears to be very doubtful.

Pair #7- Loden 1171/Loden 1194

Although there is a slight overdensity of stars near the position of Loden 1171, we have been unable to found any evidence of this OC in *Gaia* EDR3. Moreover, the reported $\mu_\alpha$ in the literature vary widely from -1.0 ± 3.7 (Dias et al. 2014) to -13.5 ± 0.3 (Zejda et al. 2012). There are also diverse reports on heliocentric distance, from 500 pc (Kharchenko et al. 2009) to 789 pc (Joshi et al. 2016), as well as on angular diameter, from 9 arcmin (Bica et al. 2019) to 17.4 arcmin (Joshi et al. 2016). Neither parallax nor RV has been reported. Thus, it may well be a mere asterism, in which case pair #7 would be illusory.

Pair #8 – ASCC 34/Loden 46

There is no evidence of ASCC 34 in Gaia EDR3, despite the presence near its alleged center (at galactic coordinates 209.679 +7.030) of a small clump of stars resembling a cluster core, which is just a chance alignment according to Gaia data. However, let's consider that ASCC 34 is a real OC, assuming that the existing literature is correct. From the data reported in DFM, the difference in $\mu_\alpha$ between both OCs of this candidate pair would be 11.75 mas/yr. At the assumed distance of this OC pair by DFM (540 pc), this would imply a differential tangential rate of more than 30 km/s, which would be too high for a binary system of OCs (Conrad et al. 2017). Moreover, the reported distances of ASCC 34 are 550 pc (Kharchenko et al. 2009) and 477 pc (Kharchenko et al. 2013), so that its average distance would be less than half the distance derived from Gaia data for Loden 46 (Table 1). Most importantly, the difference in galactic longitude of both objects (~73$^{o}$) indicates that this pair is a selection error in DFM. For either of the given reasons, this candidate pair can be ruled out.

Pair #9 – Loden 46/NGC 3228

Candidate pair #9 is formed by NGC 3228 and, again, by Loden 46, i.e., DFM is talking about a possible triple system. In this case, the existence of both OCs is undeniable, and their positions are close enough. The PMs are also (marginally) compatible. However, Loden 46 age is very poorly constrained, ranging from 0.10 Gyr (Kharchenko et al. 2013) to 1.07 Gyr (Kharchenko et al. 2009). Since both estimations come from the same group of authors, we adopt the most recent determination as most accurate. However, in that case, Loden 46 would be a young OC, and thus this pair of OCs would be out of the scope of the present analysis. Anyway, the reported RV for NGC 3228 (-22.4 km/s; Dias et al. 2002; Kharchenko et al. 2013) is incompatible with the RV for Loden 46, which ranges from 24.3 km/s (Tarricq et al. 2021) to 26.7 km/s (Dias et al. 2021). In any case, their *Gaia* derived distances and parallaxes are incongruent: NGC 3228 is much closer to the Sun than Loden 46, which discards the actual existence of the alleged binary cluster and the triple system.

Pair #10 –Ruprecht 39/NGC 6469

Again, most parameters of this candidate pair are disparate, including parallaxes, distances, $\mu_\delta$, ages and RVs (Table 1). The median reported heliocentric distance of Ruprecht 139, 0.59 kpc (Joshi et al. 2016), close to the distance given in DFM (0.55 kpc), is incongruent with either the *Gaia* EDR3 parallax obtained in the present study or the reported distance to NGC 6469. The presence of nebulosity around Ruprecht 139 is at odds with its older reported age (1.1 Gyr; Kharchenko et al. 2013). Although this age has been quoted routinely in the literature, the Kharchenko et al. (2013) catalogue is not very suitable as a source of ages for young OCs. The reason is that the listed parameters (except PMs) are based on near-IR photometry (2MASS), and the corresponding CMDs have low age sensitivity in this age interval. A much younger age has been obtained recently (0.004 Gyr; Liu & Pang 2019). If Ruprecht 139 is a young OCs after all, the inclusion of Pairs #10 and #12 in the present analysis would be pointless. Whatever the case, one of the plausible star members of Ruprecht 139 (*Gaia* EDR3 source 4069123828085540992) has RV = 68 km/s, which is incompatible with the reported RV of NGC 6469 (Table 1). All in all, the physical link of this pair is rather unlikely.

Incidentally, during the study of this region using *Gaia* EDR3, three other OCs were identified that appear to be close to Ruprecht 139 and share compatible astrometric parameters, namely LP 1625, LP 1208, and LP 1209 (Liu & Pang 2019), and could therefore be associated. However, the study of this possible group of OC is out of the scope of the present report.

Pair #11- Johansson 1/ Alesi 8

These OCs have been recently identified, via Gaia DR2, with the associations of stars Theia 353 and 335, respectively (Kounkel & Covey 2019). Theia 353 is a string of stars of mean parallax 1.27 mas, while Theia 335 has parallax 1.48 mas. The difference in parallaxes is at least 16%, and corresponds to a distance between both objects of more than 110 pc that would exclude their present physical link. In addition, some of the astrometric parameters are not well-matched. For example, the Y positions in the heliocentric XYZ reference frame are -371 pc and -561 pc for Theia 335 and 353, respectively. Therefore, the existence of a binary cluster formed by these members is doubtful, despite both ages being very similar (and very close to the limit age for the present study): 102 and 106 Myr for Theia 335 and 353, respectively. However, the possibility that this pair could be a primordial one that has relaxed to the point that it does not meet the criteria that work for younger open clusters cannot be excluded at present.

Pair #12 – Bochum 14/Ruprecht 139

The physical link between Bochum 14 and Ruprecht 139 seems likely at first sight since they are apparently close, and their parameters are at least marginally compatible (Table 1). The reported distances to Bochum 14 vary from 0.54 kpc (Joshi et al. 2016) to 0.97 kpc (Morales et al. 2013). This range is compatible with the reported distance of Ruprecht 139 but in contrast with the *Gaia* EDR3 parallaxes of both OCs, which lead to derived distances of at least 3 kpc. Bochum 14 is a young cluster embedded in its parent molecular cloud, even if there is no consensus on its exact age since reported values span from 1 Myr (Battinelli et al. 1994) to 39 Myr (Liu & Pang 2019). There is no consensus at all on the age of Ruprecht 139 (see pair #10). All in all, the gravitational capture of this pair is dubious since the gravitational link of both OCs, and the old age of Ruprecht 139 require confirmation. If Ruprecht 139 is young enough, this pair could be a primordial one.

Pair #13 - Loden 565/ASCC 68

This alleged double cluster would be formed by Loden 565 and ASCC 68 (also known as [KPR2005] 68). A detailed study based on *Gaia* DR2 (Perren et al. 2020) has concluded that Loden 565 is most likely a random stellar fluctuation. Thus, there is no case for any pair of OCs involving Loden 565. By the way, six out of sixteen OCs in the cited study were found to be mere stellar fluctuations, calling for a thorough revision of assumed OCs in the pre-*Gaia* literature. Such work has been recently undertaken by refuting 38 'well-known' OCs (Cantat-Gaudin & Anders 2020).

Pair #14- VdBH91/Ruprecht 89

Perren et al. (2020) have also concluded that VdBH 91 is a random stellar fluctuation and not a real OC. Therefore, this candidate pair is just a misconception, despite previously reported PMs and distances of both members seemed well-matched.

Pair #15-ASCC 4/NGC 189

This candidate pair is formed by ASCC 4 (or [KPR2005] 4) and NGC 189. The literature on ASCC 4 indicates that it should be detected in Gaia data since its reported distances range from 0.55 kpc (Kharchenko et al. 2013; Dib et al. 2018) to 0.75 kpc (Kharchenko et al. 2009). However, we have not found any Gaia-based study that includes it nor any trace of its existence through manual mining of Gaia EDR3. Besides, the literature PMs are assorted, and the reported RV is comparable with its associated error (Conrad et al. 2017). Our preliminary conclusion is that this cluster of stars is a chance alignment. If confirmed, no group containing ASCC 4 could exist.

Pair #16- NGC 1746/NGC 1758

After reexamining NGC 1746 using *Gaia* DR2 data, Cantat-Gaudin & Anders (2020) concluded that this alleged OC is a mere asterism. However, NGC 1750 is an apparently close OC that could be linked to NGC 1758. Moreover, sometimes all three objects have been considered one single object, catalogued as NGC 1746 (e.g., Kharchenko et al. 2013). Thus, we have reconsidered the possible pair of NGC 1750 and NGC 1758. Nevertheless, the reported RVs of -7.5 and 11 km/s, respectively (Tarricq et al. 2021), make unlikely their link. The rest of the astrometric *Gaia* measurements confirm the disparity of both OCs (Cantat-Gaudin et al. 2020). Hence, this candidate pair is, once more, not an actual binary system.

Pair #17- Basel 5/NGC 6425

Diverse distances, PMs and ages have been reported for Basel 5, but parallax and RV of this object are unknown. It is located in a very crowded field of the Milky Way. Our reexamination with *Gaia* EDR3 obtained no evidence of such OC (several halo stars probably related to NGC 6451 were recovered, instead). Accordingly, Cantat-Gaudin & Anders (2020) considered Basel 5 a mere asterism caused by extinction patterns. So, any alleged pair containing it would not exist.

Pair #20- Ruprecht 91/ESO 128-16

This candidate pair was classified as a hyperbolic encounter by DFM because of their disparate kinematics. Therefore, it is not a linked system. Moreover, ESO 128-16 could be a spurious overdensity of stars since we have not found any trace of it by manual mining of Gaia EDR3, although it has been included in the catalog of Hao et al. (2021) (see, however, comment on pair #16b).

Pair #21- NGC 2447/NGC 2448

DFM and Conrad et al. (2017) proposed this binary OC. Despite some similar parameters, their different PMs suggest that both OCs may not be a physical system (Table 1). For instance, $\mu_\delta$ are 5.1 and 2.9 mas/yr, respectively (Cantat-Gaudin et al. 2020). The combined *Gaia* data for PM and parallax led to a difference in tangential velocity up to 12 km/s, beyond the limit adopted for linked OCs. The difference in parallaxes, higher than 10%, seems to be also significant. In fact, from the most recently reported Galactocentric coordinates (Tarricq et al. 2021), both OCs would be 114 pc apart, again exceeding the 100 pc permissive limit. Accordingly, their orbital parameters significantly differ. For example, the orbital pericentres are 9.781 ± 0.035 kpc and 9.145 ± 0.181 kpc (Tarricq et al. 2021). Overall, pair #21 appears to be, at least, dubious.

There is no consensus on the age of NGC 2447. For instance, Liu & Pang (2019) report 1.15 Gyr, while others obtain 0.58 Gyr (Conrad et al. 2017; Cantat-Gaudin et al. 2020). In any case, this OC seems much older than NGC 2448 (Table 1).

Pair #22- Biurakan 2/Ruprecht 172

All reported RVs of Biurakan 2 range from -19.7 km/s (Dias et al. 2014) to -24.9 (Zhong et al. 2020), with a consensus value of -22 km/s (Vande Putte et al. 2010; Kharchenko et al. 2013; Loktin & Popova 2017; Conrad et al. 2017). This RV is at odds with the reported RVs of Ruprecht 172: 14.1 km/s (Tarricq et al. 2021) and 15.4 km/s (Soubiran et al. 2018). In addition, Gaia derived parallaxes, distances, PMs and ages are incongruent for this candidate pair (Table 1). Although the disparity in ages is irrelevant in the present scrutiny, the mismatch of the rest of the parameters strongly indicates that these OCs form an optical pair since Ruprecht 172 is much farther away than Biurakan 2.

Pair #23- NGC 6242/Trumpler 24

This candidate pair was classified as a hyperbolic encounter by DFM, and their incongruent PMs and RVs support this view (Table 1). However, Gaia's mean parallaxes indicate that Trumpler 24 is much further away than NGC 6242. In any case, this pair cannot be a binary system.

On the other hand, Trumpler 24, a young and scattered embedded cluster, seems to be associated with NGC 6231 (Yalyalieva et al. 2020) and ESO 332-8 (Dias et al. 2021), two young OCs in the same star-forming complex.

**TABLE 1.** Selected possible open cluster groups (Gr) from DFM and candidate member's properties revisited in this study (Section 3). Column headings: 1. Group number; 2. OC name; 3. Galactic longitude; 4. Galactic latitude; 5. Parallax; 6. Photometric distance; 7. PM in right ascension; 8. PM in declination; 9 OC radius; 10. Number of member stars; 11. Age; 12. Radial velocity. The suffix b in some of the groups refers to groups from the NCOVOCC catalog (Dias et al. 2002), as explained in the text. Abbreviations: [a] radius containing 50% of members; [c] maximum cluster member's distance to average position; [e] reexamined using *Gaia* EDR3 due to insufficient, imprecise or inconsistent reports; [f] too few stars for a complete characterization; [g] see text; [p] protocluster or embedded cluster; + et al.

| ***1*** | ***2*** | ***3*** | ***4*** | ***5*** | ***6*** | ***7*** | ***8*** | ***9*** | ***10*** | ***11*** | ***12*** | ***13*** |
|---|---|---|---|---|---|---|---|---|---|---|---|---|
| Gr | OC | *l* | *b* | *plx* | *d* | $\mu_\alpha$ | $\mu_\delta$ | *R* | *N* | *Age* | RV | References and notes |
| | Name | degree | degree | mas | kpc | mas yr$^{-1}$ | mas yr$^{-1}$ | arcmin | stars | Gyr | Km/s | |
| 1 | ASCC 101 [KPR2005] 101 | 68.03 | 11.61 | $2.4_9$ | 0.41 | $0.9_3$ | $1.2_9$ | 22.3[a] | 69 | 0.49-0.33[g] | -15 - -32[g] | Cantat-Gaudin+ 2020 |
| 1 | ASCC 100 [KPR2005] 100 | 64.4 | 12.67 | 2.9 ± 0.4 | | 2.2 ± 0.3 | -3.1 ±0.3 | 25 | 24 | $0.08_9$-$0.10_2$[g] | $-22._9$ - $-25._9$[g] | This work[e] |
| 5 | ASCC 90 [KPR2005] 90 | 354.22 | -1.95 | $1.7_1$ | 0.56 | $-1.6_3$ | $-2.6_8$ | 23.3[a] | 58 | 0.81-0.65[g] | 8 | Cantat-Gaudin+ 2020 Tarricq et al. 2021 |
| 5 | NGC 6405 | 356.58 | -0.76 | $2.1_7$ | 0.46 | $-1.3_1$ | $-5.8_4$ | 16.5[a] | 573 | $0.03_4$ | | Cantat-Gaudin+ 2020 |
| | | | | | | | | | | | -9[g] | |
| 8 | ASCC 34 | 209.67 | 7.02 | - | 0.48- | -2.1 | -0.4 | 12 | 22 | 0.34 | - | Kharchenko+ 2013 |
| | [KPR2005] | | | | 0.55[g] | | | | | | | |
| | 34 | | | | | | | | | | | |
| 8 | Loden 46 | 282.56 | 2.25 | 0.88 | 1.1 | -11.4 | 4.6 | 20[a] | 44 | 0.10- | 24 - | Cantat-Gaudin+ 2018 |
| 9 | | | | | | | | | | $1.0_7$[g] | 27[g] | |
| 9 | NGC 3228 | 280.76 | 4.49 | $2.0_4$ | 0.50 | -14.9 | -0.7 | $30._7$[a] | 117 | 0.03 | -22[g] | Cantat-Gaudin+ 2020 |
| 10 | NGC 6469 | 6.56 | 1.97 | 0.59 | 1.7 | 0.6 | 0.8 | 3.7[a] | 48 | 0.07 | | Cantat-Gaudin+ 2020 |
| | | | | | | | | | | | $-7._3$ | Tarricq et al. 2021 |
| 10 | Ruprecht | 6.43 | -0.18 | 0.29 ± 0.06 | | 0.0 ± 0.2 | -1.4 ±0.2 | 7 | 65 | $0.00_4$- | 68±3 | This work[e] |
| 12 | 139 | | | | 0.59 | | | | | $1.1_2$[g] | | Joshi et al. 2016 |
| 12 | Bochum 14 | 6.38 | -0.50 | 0.32 ± 0.06 | 0.54- | 0.3 ± 0.3 | -1.2 ±0.2 | 2 | 44 | $0.00_1$- | - | This work[e] |
| | | | | | 0.97[g] | | | | | $0.03_7$ | | |
| 21 | NGC 2447 | 240.05 | 0.15 | 0.97 | 1.0 | -3.8 | 3.9 | 12[a] | 731 | 0.58 | 22 | Conrad et al. 2017 |
| | | 240.04 | 0.15 | | 1.0 | -3.6 | 5.1 | | | 0.58 | | Cantat-Gaudin+ 2020 |
| 21 | NGC 2448 | 240.76 | -0.26 | 0.88 | 1.0 | -3.8 | 4.7 | 16[a] | 121 | 0.02 | 24 | Conrad et al. 2017 |
| | | 240.85 | -0.43 | | 1.1 | -3.4 | 2.9 | | | 0.10 | | Cantat-Gaudin+ 2020 |
| 22 | Biurakan 2 | 72.75 | 1.36 | 0.54 | 1.7 | -3.2 | -6.8 | 8.3[a] | 47 | $0.00_9$ | -20 - | Cantat-Gaudin+ 2020 |
| | | | | | | | | | | | -25[g] | |
| 22 | Ruprecht | 73.11 | 1.01 | 0.26 | 3.6 | -2.0 | -3.7 | 2.5[a] | 82 | $1.0_5$ | 14 - | Cantat-Gaudin+ 2020 |
| | 172 | | | | | | | | | | 15[g] | |
| 23 | NGC 6242 | 345.45 | 2.46 | 0.76 | $1.2_4$ | 1.1 | -0.8 | 6.2[a] | 471 | $0.07_8$ | | Cantat-Gaudin+ 2020 |
| | | | | | $1.2_1$ | | | | | $0.08_3$ | 2 | Tarricq et al. 2021 |
| 23 | Trumpler 24 | 344.63 | 1.59 | 0.59 | | $-0.2_7$ | -1.3 | | 327 | $0.00_8$ | -35 | Hao et al. 2021 |
| 27 | Basel 8 | 203.85 | -0.16 | 0.63 | 1.5 | -0.1 | -2.4 | 19[a] | 14 | 0.05- | | Cantat-Gaudin+ 2020 |
| | | | | | | | | | | 0.13[g] | 11 | Conrad et al. 2017 |
| 27 | NGC 2251 | 203.61 | 0.11 | 0.66 | 1.5 | 0.7 | -3.8 | 7.4[a] | 109 | 0.28 | | Cantat-Gaudin+ 2020 |
| | | | | | 1.5 | | | | | 0.28 | 26 | Tarricq et al. 2021 |
| 33 | Ruprecht | 233.08 | 3.24 | 0.87 | $1.1_3$ | -4.3 | 3.2 | 6.8[a] | 41 | 0.45 | | Cantat-Gaudin+ 2020 |
| | 151 | | | | $1.0_9$ | | | | | 0.49 | 26 | Tarricq et al. 2021 |
| 33 | NGC 2428 | 233.09 | 2.70 | 0.71 | $1.3_2$ | -3.3 | 2.5 | 8.8[a] | 163 | 0.72 | | Cantat-Gaudin+ 2020 |
| | | | | | $1.2_9$ | | | | | 0.74 | 53 | Tarricq et al. 2021 |
| 9b | Loden 46 | 282.56 | 2.25 | 0.88 | 1.1 | -11.4 | 4.6 | 20[a] | 44 | 0.10- | - | Cantat-Gaudin+ 2018 |
| | | | | | | | | | | $1.0_7$[g] | | |
| 9b | ASCC 59 | 283.83 | -0.76 | 0.30 ± 0.03 | | -5.1 ± 0.2 | 3.6 ± 0.2 | 18 | 64 | | -4 | This work[e] |
| | [KPR2005] | 283.82 | -0.52 | - | $0.5_1$ | -4.8 | 3.2 | 20 | 219 | 0.29 | - | Kharchenko+ 2013 |
| | 59 | | | | | | | | | | | |
| 23b | Pismis 19 | 314.71 | -0.31 | 0.26 | 3.5 | -5.5 | -3.2 | 2.1[a] | 430 | 0.63- | | Cantat-Gaudin+ 2020 |
| | | | | | | | | | | $1.1_2$[g] | $-29._6$ | Tarricq et al. 2021 |
| 23b | Trumpler 22 | 314.66 | -0.59 | 0.39 | 2.4 | -5.1 | -2.7 | 6.2[a] | 140 | $0.02_4$- | -38 - | Cantat-Gaudin+ 2020 |
| | | | | | 2.4 | | | | | 0.31[g] | -43[g] | Tarricq et al. 2021 |
| 26b | NGC 2421 | 236.28 | 0.06 | $0.34_3$ | $2.6_8$ | -3.1 | 3.1 | 5.8[a] | 406 | 0.07- | | Cantat-Gaudin+ 2020 |
| | | | | | $2.5_8$ | | | | | 0.09[g] | 89 | Tarricq et al. 2021 |
| 26b | Czernik 31 | 236.27 | 0.27 | $0.29_5$ | 3.4 | -1.9 | 3.0 | 2.3[a] | 71 | 0.02- | | Cantat-Gaudin+ 2020 |
| | | | | | 3.2 | | | | | 0.18[g] | 103 | Tarricq et al. 2021 |

Pair #25- ASCC 6/Stock 4

Although this candidate pair is included in the present revision because, according to DFM, both OCs would be older than 100 Myr, the most recent reports disagree on the ages. For ASCC 6 ([KPR2005] 6), the two most recent determinations, based on *Gaia* data, agree on the age of ca. 0.04 Gyr (Bossini et al. 2019; Cantat-Gaudin et al. 2020). For Stock 4, the *Gaia* derived ages are ca. 0.07 Gyr (Cantat-Gaudin et al. 2020; Tarricq et al. 2021). Thus, this pair may well be a primordial system of very young OCs.

Pair #27- Basel 8/NGC 2251

The reported ages of Basel 8 span from 0.048 Gyr (e.g., Conrad et al. 2017) to 0.13 Gyr (Loktin & Popova 2017). Whatever the case, NGC 2251 seems to be older (Table 1), which motivates the inclusion of pair #27 in this revision. Although the celestial positions and distances for Basel 8 and NGC 2251 are well-matched, their PMs are not (Table 1). From PMs and parallaxes, a difference in tangential velocity of 12 km/s, higher than the agreed threshold, is obtained. The divergent RVs also cast doubt on the physical link of this pair of OCs.

Pair #31- Loden 165/Carraro 1

A reexamination of the star field of Loden 165 with the *Gaia* EDR3 catalogue did not show any trace of this OC, or at least no OC compatible with the literature data of Loden 165 (some star members of VdBH 99 and NGC 3293 were found, instead). Accordingly, there is no consensus on the reported distances, PMs and ages of this star cluster. Moreover, there is no data on its parallax or RV. Thus, we suggest that Loden 165 is just an asterism, and there is no pair of OCs including it.

Surprisingly enough, one of the reported distances and one of the reported ages of Loden 165 and its alleged companion Carraro 1 exactly coincide. Both OCs would be at 1900 pc from us, and their log(age in yr) would be 9.48 (3.02 Gyr). When looking for the source of these particular data, we reached in both cases the WEBDA catalogue. This catalogue provides the bibliographic reference of the data for Loden 165 (Carraro et al. 2001). However, it does not include any data details or bibliographic sources for the listed age and distance of Carraro 1. Perhaps these coincident data may be due to a mistake favoured by the close position of both objects, and perhaps all these coincidences led to the erroneous assumption that both OCs form a double system.

Pair #33- Ruprecht 151/NGC 2428

The astrometric *Gaia* data on Ruprecht 151 and NGC 2428 do not coincide. Both parallaxes and photometric distances indicate that Ruprecht 151 is ca. 200 pc closer to us than NGC 2428 (Table 1). PMs and RVs confirm the mismatch. As a result, the orbital parameters are also significantly different (Tarricq et al. 2021). Therefore, this is merely an optical pair.

### *3.2. NCOVOCC Catalog*

The numbers of candidate pairs based on this catalogue data are noted #1b, #2b, and so on to differentiate them from the candidate pairs from the WEBDA database.

Pairs #2b, #6b, #8b, #10b, #11b, #12b, #15b, #17b, #21b and #25b coincide with WEBDA pairs #1, #5, #9, #11, #15, #17, #21, #22, #27 and #31, respectively, which have already been discussed in sub-section 3.1.Pair #9b- Loden 46/ASCC 59

We find, again, many disparities between both alleged members. The previously reported data on ASCC 59 were incomplete and somewhat inconsistent. Thus, we decided to reexamine it using Gaia EDR3 (Table 1). However, the new parallax is at odds with the reported distances of ASCC 59. Most importantly, the Gaia data on Loden 46 (see pairs #8 and #9) does not match any of the data of ASCC 59. Therefore, candidate pair #9b can also be surely discarded.

Pair #16b- ESO 132-14/ NGC 5281

Despite the existence of a dense clump of a few stars at the centre of ESO 132-14 that resembles an OC core, our manual inspection of *Gaia* EDR3 shows no trace of such an OC and reveals that the apparent core is a chance alignment of stars. However, *Gaia* should detect that cluster since it was reported to be only 1.1 kpc away (Dias et al. 2002; Morales et al. 2013; Dib et al. 2018). On the other hand, the catalogue of Hao et al. (2021) reports a mean parallax of 0.39 mas, at odds with that distance of ESO 132-14. Parenthetically, this all-inclusive catalogue (3794 entries) encompasses numerous asterisms (e.g., NGC 1663, NGC 1746, Ruprecht 46, Ruprecht 155, Collinder 471, Basel 5, Loden 1 (Cantat-Gaudin & Anders 2020)) and duplicated OCs with different listed parameters (e.g., Alessi 44, An-drews-Lindsay 5, Arp-Madore 2, Havlen Moffat 1,VdBH 121), especially those in the ESO series (e.g., ESO 021-06, ESO 313-03, ESO 313-11, ESO 332-13, ESO 334-02, ESO 368-11, ESO 368-14, ESO 392-13, ESO 397-01, ESO 429-02) . In addition, the previous literature on ESO 132-14 shows assorted PMs (e.g., Kharchenko et al. 2013; Dias et al. 2014) and neither RV nor mean parallax measurements. Given all that conflicting evidence, we consider this OC and the candidate pair unlikely to be physical.

Pair #23b- Pismis 19/Trumpler 22

The age of Trumpler 22 is poorly constrained. Pre-*Gaia* studies report ages up to 0.31 Gyr (e.g., Kharchenko et al. 2013), but most recent works report ages close to 0.03 Gyr (Bossini et al. 2019; Cantat-Gaudin et al. 2020; Tarricq et al. 2021). Anyway, these values do not overlap with the reported ages of Pismis 19, a redder and older cluster with ages that range from 0.63 Gyr (Buckner & Froebrich 2014) to 1.12 Gyr (Morales et al. 2013). Such age disparity prompted the inclusion of this candidate pair in the present study.

Even if positions and PMs are compatible, distances and parallaxes are not. Pre-*Gaia* catalogs assigned to Trumpler 22 photometric distances as low as 1.5 kpc (e.g., Dias et al. 2002). However, *Gaia* DR2-based studies agree on a photometric distance of 2.4 kpc (Table 1), which harmonizes with the measured parallax allowing for the mentioned global offset of *Gaia* DR2 parallaxes (Lindegren et al. 2018). In any case, all reported distances are significantly smaller than the distance (and the corresponding parallax) of Pismis 19 (Table 1).

The reported RVs are not enlightening in this case. For Trumpler 22, RVs span from -38 km/s (Dias et al. 2002; Loktin & Popova 2017) to -43 km/s (Soubiran et al. 2018; Tarricq et al. 2021), which are only marginally compatible with the reported RV for Pismis 19 (Table 1). Anyway, the ensemble of data suggests that this pair is a chance alignment, mainly because Trumpler 22 is more than one kpc closer to the Sun than Pismis 19.

On the other hand, the younger OC Trumpler 22 has been physically connected to NGC 5617 (De Silva et al. 2015). According to this study, the two close OCs share similar age (~70 Myr), average RV (-38.5 ± 2.0 km/s), and mean metallicity (−0.18 ± 0.02 dex), forming a primordial binary cluster.

Pair #26b- NGC 2421/Czernik 31

NGC 2421 is a well-known OC, but its age was poorly constrained before *Gaia*. However, four of the five *Gaia*-based studies converge on ages ranging from 0.066 Gyr (Cantat-Gaudin et al. 2020) to 0.091 Gyr (Tarricq et al. 2021). The age of Czernik 31 was also poorly constrained: reports range from 0.021 Gyr (e.g. Bossini et al. 2019) to 0.18 Gyr (Kharchenko et al. 2013). The latter tentative age (see comment on pair #10), adopted in DFM, motivates the inclusion of this pair in the present work. Anyway, inspection of the data summarized in Table 1 suggests that this is an optical pair. Most of the parameters are incompatible. In particular, both the photometric distances and the mean Gaia parallaxes indicate that NGC 2421 is significantly closer than Czernik 31.The reported RVs and $\mu_\alpha$ also do not match.

None of the revised candidate pairs is retrieved in the more recent and constraining study of Conrad et al. (2017), except for the questionable case of NGC 2447 and NGC 2448. Moreover, Conrad et al. (2017) warn that possibly many of the previously proposed groupings in the literature were not recovered in their survey because they are not real.

In sum, we have not found any likely binary clusters from the DFM study with any of their members older than 100 Myr. Some of the pairs are optical pairs, others are hyperbolic encounters, and a few pairs may be primordial pairs with flawed ages. A significant number of the clusters studied in this sample are most likely false OCs, namely Loden 1171, ASCC 34, Loden 565, VdBH 91, ASCC 4, NGC 1746, Basel 5, Loden 165, ESO 128-16 and ESO 132-14. The higher frequency of Loden objects suggests that that series of OCs might contain more flawed OCs than average. The preliminary conclusion of this section is that our working hypothesis has passed this trial and that most double or multiple OCs are primordial groups.

**TABLE 2.** Selected possible open cluster groups (Gr) and candidate member's properties studied in this work (Sections 4 and 5). Column headings: 1. Group label; 2. OC name; 3. Galactic longitude; 4. Galactic latitude; 5. Parallax; 6. Photometric distance; 7. PM in right ascension; 8. PM in declination; 9 OC radius; 10. Number of member stars; 11. Age; 12. Radial velocity. Abbreviations: [a] radius containing 50% of members; [c] maximum cluster member's distance to average position; [e] reexamined using *Gaia* EDR3 due to insufficient, imprecise or inconsistent reports; [f] too few stars for a complete characterization; [g] see text; [p] protocluster or embedded cluster; + et al; (?) unlikely group.

| *1* | *2* | *3* | *4* | *5* | *6* | *7* | *8* | *9* | *10* | *11* | *12* | *13* |
|---|---|---|---|---|---|---|---|---|---|---|---|---|
| Gr | OC<br>Name | $l$<br>degree | $b$<br>degree | *plx*<br>mas | *d*<br>kpc | $\mu_\alpha$<br>mas yr$^{-1}$ | $\mu_\delta$<br>mas yr$^{-1}$ | *R*<br>arcmin | *N*<br>stars | *Age*<br>Gyr | RV<br>Km/s | References and notes |
| - | SAI 25 | 139.70 | -1.33 | 0.30 ± 0.11 | 1.1-<br>2.7[g] | 0.2 ± 0.2 | $-0.7_5$ ± 0.3 | 2.5 ±<br>0.5 | 19[f] | 1.5-<br>$0.00_2$[g] | -<br>-66 | This work[e]<br>Tarricq et al. 2021 |
| A | NGC 6823 | 59.42 | -0.14 | 0.45 | 2.3<br>2.2 | -1.7 | -5.3 | 4.4[a] | 140 | $0.00_2$-<br>$0.01_0$[g] | 11-<br>30[g] | Cantat-Gaudin+ 2020<br>Tarricq et al. 2021 |
| A | Roslund 2 | 60.21 | -0.18 | 0.46 | 2.1<br>2.0 | -1.7 | -5.1 | 9.6[a] | 97 | $0.00_6$-<br>$0.01_2$[g] | <br>127 | Cantat-Gaudin+ 2020<br>Tarricq et al. 2021 |
| B | Hogg 15 | 302.05 | -0.24 | 0.27 | 3.0<br>2.9 | -6.0 | -0.5 | 3.8[a] | 72 | $0.02_0$-<br>$0.00_2$[g] | <br>-23 | Cantat-Gaudin+ 2020<br>Tarricq et al. 2021 |
| B | La Serena 39 | 301.10 | -0.17 | 0.30 ± 0.08 | | -5.9 ± 0.3 | -0.4 ± 0.2 | 3.0 ±<br>0.5 | 16[f] | | - | This work[e] |
| B | [MCM2005 b]32 | $300.13_5$ | $-0.08_5$ | 0.26 ± 0.07 | | -6.1 ± 0.2 | -0.3 ± 0.2 | 1.0 ±<br>0.5 | 15[f] | | -<br>-39 | This work[e]<br>Morales et al. 2013 |
| C | UBC 344 | 18.35 | 1.82 | 0.47 | $1.9_2$<br>$1.8_4$ | -0.3 | -2.2 | 9.7[a] | 314 | $0.00_3$<br>$0.00_3$ | <br>42 | Cantat-Gaudin+ 2020<br>Tarricq et al. 2021 |
| C | NGC 6604 | 18.24 | 1.69 | 0.45 | $1.8_9$ | $-0.4_5$ | -2.3 | 1.0-<br>17[g] | 88 | $0.00_3$-<br>$0.00_8$[g] | -5 –<br>136[g] | Dias et al. 2021 |
| C | [BDS2003] 9 | 18.67 | 1.97 | 0.49 ± 0.10 | | -0.5 ± 0.3 | -2.0 ±0.3 | 1.5 ± 0.5 | 16[f] | | - | This work[e p] |
| C | Casado 67 | 18.78 | 1.83 | 0.47 ± 0.08 | 2.2 | $-0.4_5$ ± 0.2 | $-2.0_5$ ±0.3 | 2.5 ± 0.5 | 27 | 0.01 | - | This work[p] |
| D | NGC 6383 | 355.67 | 0.06 | 0.87 | $1.1_2$ | 2.6 | -1.7 | 4.9[a] | 245 | $0.00_4$ | | Cantat-Gaudin+ 2020 |
| | | | | | $1.0_9$ | | | | | $0.00_4$ | 2 | Tarricq et al. 2021 |
| D | Casado 68 | 354.54 | 0.20 | 0.87 ± 0.10 | 1.2 | $2.5_5$ ± 0.4 | -1.8 ± 0.4 | 13 ± 1 | 52 | 0.01 | 4 | This work[p] |
| E | NGC 1893 | 173.58 | -1.63 | 0.27 | $3.2_2$ | -0.2 | -1.4 | 5.1[a] | 123 | $0.00_4$ | | Cantat-Gaudin+ 2020 |
| | | | | | $3.1_5$ | | | | | $0.00_4$ | -4 | Tarricq et al. 2021 |
| E | Casado 69 | 173.16 | -1.30 | 0.30 ± 0.09 | | -0.3 ± 0.3 | -1.6 ± 0.4 | 3.5±0.5 | 17[f] | | - | This work[p] |
| F | NGC 6193 | 336.69 | -1.58 | 0.81 | $1.2_6$ | 1.3 | -4.1 | 9.4[a] | 428 | $0.00_5$ | | Cantat-Gaudin+ 2020 |
| | | | | | $1.2_3$ | | | | | $0.00_5$ | -76 | Tarricq et al. 2021 |
| F | Casado 70 | 336.35 | -1.20 | 0.84 ± 0.09 | 1.1 | 1.3 ± 0.5 | -4.3 ± 0.4 | 7 ± 1 | 63 | 0.01 | - | This work [p] |
| G | NGC 3572 | 290.74 | 0.17 | 0.38 | $2.4_6$ | -6.3 | 1.9 | 4.4[a] | 75 | $0.00_5$ | | Cantat-Gaudin+ 2020 |
| | | | | | $2.3_5$ | | | | | $0.00_5$ | 1 | Tarricq et al. 2021 |
| G | Hogg 10 | 290.78 | 0.15 | 0.37 | 1.8-2.5[g] | -6.2 | 1.9 | 1-8 | 44 | $0.00_6$ | 1 - -7[g] | Hao et al. 2021 |
| G | LP 1531 | 291.02 | $0.04_4$ | $0.38_5$ | | -6.2 | 1.8 | 6.7[c] | 57 | $0.00_4$ | - | Liu & Pang 2019 |
| H | FSR 0198 | 72.18 | 2.61 | 0.49 | $2.1_8$ | -3.6 | -6.6 | 7.6[a] | 82 | $0.00_5$ | | Cantat-Gaudin+ 2020 |
| | | | | | $2.0_4$ | | | | | $0.00_5$ | 12 | Tarricq et al. 2021 |
| H | Teutsch 8 | 71.86 | 2.42 | 0.49 | $1.9_8$ | -3.5 | -6.7 | 0.4[a] | 28 | $0.00_4$ | - | Cantat-Gaudin+ 2020 |
| I | NGC 2362 | 238.18 | -5.55 | 0.74 | $1.3_4$ | -2.8 | 3.0 | 3.1[a] | 144 | $0.00_6$ | | Cantat-Gaudin+ 2020 |
| | | | | | $1.3_0$ | | | | | $0.00_6$ | 29 | Tarricq et al. 2021 |
| I | Camargo 997 | 237.50 | -4.89 | 0.80 ± 0.13 | | -2.8 ± 0.3 | 3.0 ± 0.3 | 8.5 ± 0.5 | 49 | | 30 | This work [e p] |
| J | UBC 438 | 195.70 | 0.03 | 0.21 | 3.6 | 0.2 | -0.4 | 5.8[a] | 24 | $0.00_4$ | | Cantat-Gaudin+ 2020 |
| | | | | | 3.3 | | | | | $0.00_6$ | 39 | Tarricq et al. 2021 |
| J | Casado 71 | 195.29 | 0.45 | 0.21 ± 0.06 | | 0.3 ± 0.2 | -0.5 ± 0.2 | 3.0 ± 0.5 | 14[f] | | - | This work |
| $H_1$ | NGC 6871 | 72.66 | 2.01 | 0.51 | $1.7_2$ | $-3.1_3$ | $-6.4_4$ | 22[a] | 430 | $0.00_5$ | | Cantat-Gaudin+ 2020 |
| | | | | | $1.6_6$ | | | | | $0.00_6$ | 15 | Tarricq et al. 2021 |
| $H_1$ | Teutsch 8 | 71.86 | 2.42 | 0.49 | $1.9_8$ | $-3.4_9$ | $-6.7_0$ | 0.4[a] | 28 | $0.00_4$ | - | Cantat-Gaudin+ 2020 |
| K | IC 1805 | 134.73 | $0.94_5$ | 0.45 | $1.9_6$ | -0.7 | $-0.6_7$ | 6.7[a] | 106 | $0.00_8$ | | Cantat-Gaudin+ 2020 |
| | | | | | $1.8_7$ | | | | | $0.00_7$ | -44 | Tarricq et al. 2021 |
| K | Berkeley 65 | 135.84 | 0.26 | 0.44 | $2.2_8$ | -0.7 | $-0.3_4$ | 1.5[a] | 37 | $0.09_7$ | - | Cantat-Gaudin+ 2020 |
| K | Camargo 755 | 134.81 | 1.32 | 0.45 ± 0.11 | | -0.7 ± 0.3 | -0.3 ± 0.2 | 5 ± 1 | 44 | | - | This work[e p] |
| L | VdBH 205 | 344.63 | 1.63 | 0.57 | $1.6_0$ | -0.2 | -1.1 | 5.8[a] | 55 | $0.00_6$ | | Cantat-Gaudin+ 2020 |
| | | | | | $1.5_4$ | | | | | $0.00_7$ | -2 | Tarricq et al. 2021 |
| L | ESO 332-08 | 344.39 | 1.79 | 0.53 | $1.6_9$ | -0.3 | -1.3 | | 201 | $0.00_8$ | - | Dias et al. 2021 |
| L | [DBS2003] 114 | 345.32 | 1.46 | 0.56 ± 0.10 | | -0.2 ± 0.4 | -1.4 ± 0.3 | 1.5 ± 0.5 | 19[f] | | - | This work[e p] |
| M | Berkeley 36 | 227.50 | -0.56 | $0.20_6$ | $4.3_6$ | $-1.7_3$ | $0.8_6$ | 2.8[a] | 150 | 6.8 | | Cantat-Gaudin+ 2020 |
| (?) | | | | | $4.1_3$ | | | | | 6.8 | 63 | Tarricq et al. 2021 |
| M | Casado 72 | 227.25 | -0.82 | 0.26 ± 0.06 | 3.8 | $-1.5_5$ ± | $0.8_5$ ± 0.2 | 5 ± 0.5 | 19 | 0.02 | - | This work |
| (?) | | | | | | 0.2 | | | | | | |
| N | Kronberger | 95.27 | 2.07 | 0.23 | - | $-2.6_1$ | $-3.3_2$ | 2.7[a] | 28[f] | - | | Cantat-Gaudin+ 2020 |
| (?) | 81 | | | | 4.1 | | | | | 5.6 | -85 | Tarricq et al. 2021 |
| N | Teutsch 17 | 95.31 | 1.06 | $0.25_3$ | 2.9 | $-2.8_8$ | $-3.2_1$ | | 213 | $0.01_2$- | - | Dias et al. 2021 |
| (?) | | | | $0.26_7$ | | $-2.8_8$ | $-3.2_2$ | $11._6$[c] | 58 | $0.08_3$ | | Liu & Pang 2019 |
| - | NGC 2383 | 235.27 | -2.46 | 0.28 | 3.5 | -1.6 | 1.9 | 2.2[a] | 242 | 0.26 | 55- | Cantat-Gaudin+ 2020 |
| | | | | | | | | | | | 72[g] | |
| O | NGC 2384 | 235.39 | -2.39 | 0.38 ± 0.09 | 2.0- | -2.3 ± 0.3 | 3.1 ± 0.3 | 4.0 ± | 36 | $0.00_6$- | 46- | This work[e] |
| | | | | | 3.2[g] | | | 0.5 | | $0.02_1$[g] | 53[g] | |
| O | Casado 73 | 234.69 | -2.19 | 0.33 ± 0.06 | 3.1 | -2.2 ± 0.2 | 3.1 ± 0.3 | 3.5 ± | 23 | 0.08 | - | This work |
| | | | | | | | | 0.5 | | | | |

## 4. Grouping around very young OCs

Although not necessarily so, the *Primordial group* hypothesis suggests that very young OCs (age < 0.01 Gyr) may still be associated with their siblings, i.e., those clusters born recently from the same giant molecular cloud. We have used the updated catalogue of Tarricq et al. (2021) to check out this possibility. This comprehensive catalogue lists well-studied OCs having both age and 3D kinematics. We have studied the field of those OCs to look for associated siblings using the methodology described in section 2.

### SAI 25

Since there are inconsistent reports (even from *Gaia* data) on this faint, poorly populated OC, we contribute our parameters based on *Gaia* EDR3 (Table 2). The reported distances range from 1.1 kpc (Loktin & Popova 2017) to 2.7 kpc (Cantat-Gaudin et al. 2020). Only the larger values are (marginally) compatible with *Gaia* parallaxes. Moreover, there is a much wide span of reported ages for SAI 25: from 0.002 Gyr (Cantat-Gaudin 2020; Tarricq et al. 2021) to 1.5 Gyr (e.g., Dias et al. 2002; Kharchenko et al. 2013; Hao et al. 2021). Note that only in the case of SAI 25 was indeed young, it would be justified to include it in the present section. The broad and ill-defined CMD (Cantat-Gaudin et al. 2020) has a short main sequence spanning only four magnitudes, suggesting that it is not so young. Whatever the actual age is, no associated OC has been found in a radius of 100 pc around SAI 25.

### NGC 6823 (Group A)

NGC 6823 is so young that it is still associated with the HII region LBN 059.38-00.15. Reported ages range from 2 Myr (Tarricq et al. 2021) to 10 Myr (Kharchenko et al. 2013). It also seems to be associated with Roslund 2 (Table 2), which reported ages range from 6 Myr (Kharchenko et al. 2013; Joshi et al. 2016; Dib et al. 2018) to 12 Myr (Tarricq et al. 2021). Accordingly, most characteristics of both OCs are well-matched. Unfortunately, published RVs do not allow confirming this possibility, mainly because of the wide dis-parity in Roslund 2 RVs, ranging from -5 km/s (e.g., Kharchenko et al. 2013) to 127 km/s (Soubiran et al. 2018; Tarricq et al. 2021). Notice, however, that RVs of NGC 6823 range from 11 km/s to 30 km/s so that both intervals are compatible. All in all, this pair is considered a likely double cluster candidate.

Hogg 15 (Group B)

Hogg 15 is perhaps associated with the EC La Serena 39 (Bica et al. 2019). We have reexamined this last cluster with *Gaia* EDR3, but the results are insufficient for a complete characterization due to the few stars members and the high associated errors (Table 2). RV and age are unknown, but since it is an EC, it should be very young. In the case of Hogg 15, the reported ages range from 2 Myr (Cantat-Gaudin et al. 2020; Tarricq et al. 2021) to 20 Myr (Morales et al. 2013). Anyway, the astrometric and PM data are just compatible. All in all, this candidate pair is considered dubious.

The small OC [MCM2005b]32, which has also been classified as an EC (Bica et al. 2019) and is associated with mid-IR extended emission (Mercer et al. 2005), may belong to the same group. We have performed its study using *Gaia* EDR3 since very few data on this cluster have been reported (Table 2). The RV from associated gas line velocity is barely compatible with the RV of Hogg 15 (Table 2). Although the distance between this cluster and Hogg 15 appears to be 0.13 kpc, *Gaia* EDR3 results are well-matched with those of Hogg 15. Its membership is therefore unsure. It is noteworthy at this point that if two clusters born together are on their way to separate, sooner or later, they will go beyond the 0.1 kpc limit, which is somehow arbitrary (Casado, 2021). In other words, the separation of sibling OCs may well increase with time as the systems become more and more dynamically relaxed.

Lynga 6

No close OC to Lynga 6 has been found considering both its 3D position and 3D kinematics.

UBC 344 (Group C)

According to data from Cantat-Gaudin et al. (2020), UBC 344 is a big (up to 40 arcmin) and elongated cluster (Fig. 1), containing a few cores in a nebular area. From the plot of its member stars, it became clear that one of the cores corresponds to NGC 6604, but it is unclear if NGC 6604 is part of UBC 344 or if they are two associated OCs. Note, however, that some of the UBC clusters recently discovered by Castro-Ginard et al. (2020) were already known OCs (Monteiro et al. 2020). The apparent diameter reported for NGC 6604 in the pre-Gaia literature ranges from 2.0 arcmin (Battinelli et al. 1994) to 9.0 arcmin (Bica et al. 2019). However, Liu & Pang (2019) estimate the diameter to be 34 arcmin from Gaia DR2, similar to the size of UBC 344. The centers of both objects are ca. 10 arcmin apart. Most reported ages of NGC 6604 range from 3 Myr (Dambis 1999) to 8 Myr (Kharchenko et al. 2013), and there is a broad consensus around 6 Myr (e.g., Dias et al. 2002; Ahumada & Lapasset 2007; Wu et al. 2009; Vande Putte et al. 2010; Gozha et al. 2012; Dias et al. 2021). Therefore, both clusters are very young (Table 2). The rest of the relevant parameters are also well-matched (Table 2), except for the RV of NGC 6604, which is very poorly constrained: from -5 km/s (Dias et al. 2002) to 136 km/s (Hao et al. 2021). All in all, there is a likely relationship between both objects. The general appearance resembles a rich OC in the process of disintegration (Fig. 1).

Two extra OC have been identified as possible members of Group C: [BDS 2003] 9, a cluster still embedded in the parent nebula Gum 85, and the new OC Casado 67, found in this survey (Fig. 1 and Table 2). This last object, embedded in the same giant molecular cloud, is in an area containing numerous young stellar object candidates, which suggests that it is also very young. Accordingly, its estimated age from isochrone fitting (Fig. 2) is 0.01 Gyr, compatible with the ages of other members of Group C. The derived photo-metric distance is 2.2 kpc, in good agreement with the mean *Gaia* EDR3 parallax. The estimated extinction of this new OC is $A_V$ = 4.5 mag. A small clump of ca. ten stars at galactic coordinates 18.18, 2.02 also seems to belong to the same star-forming complex.

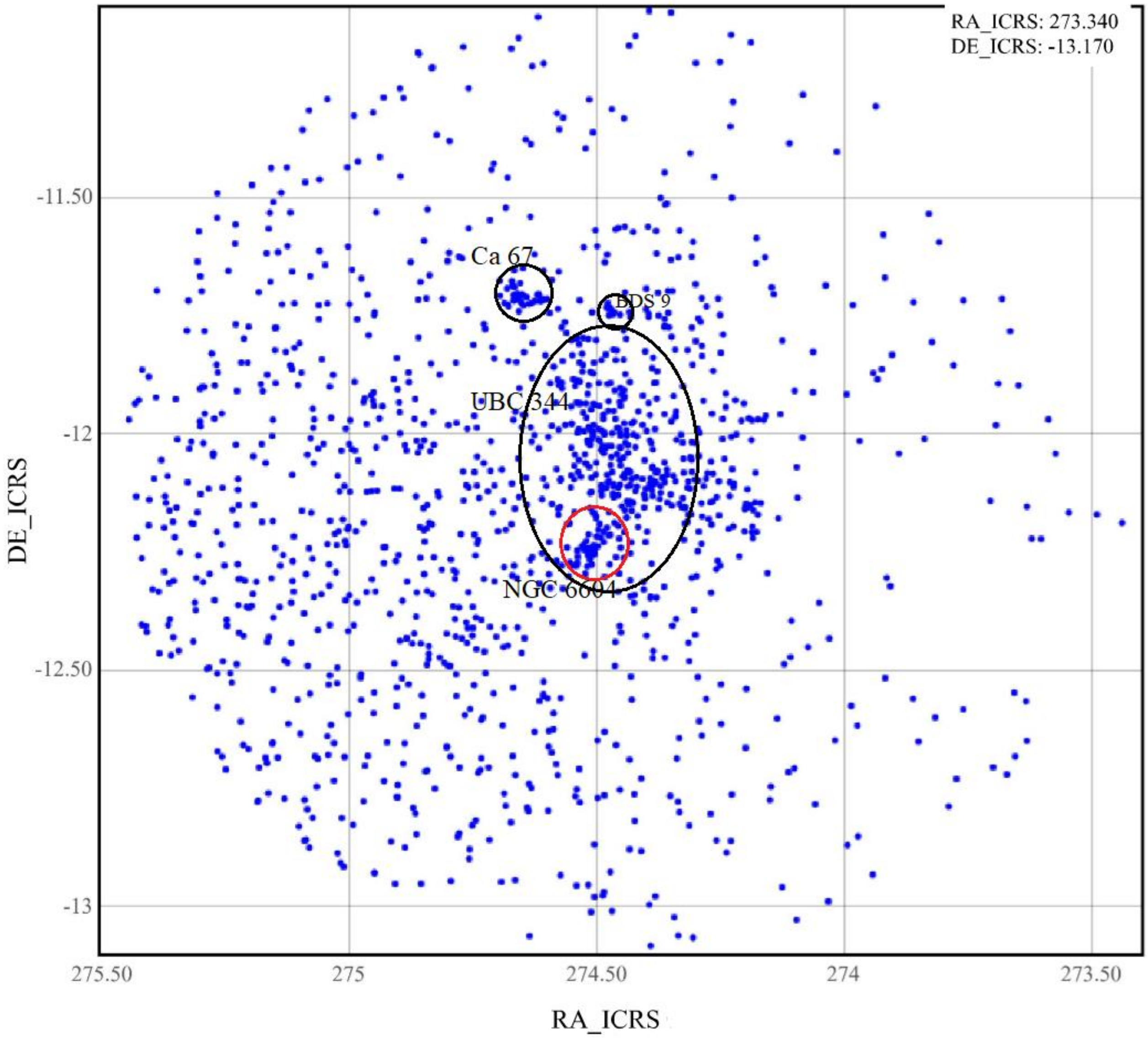

**Figure 1.** Chart of selected *Gaia* EDR3 sources defining group C. NGC 6604 is circled in red to differentiate it from the rest of UBC 344, outlined by a black ellipse. Constraints: *plx* 0.37 to 0.57 mas; $\mu_\alpha$ -0.1 to -0.9 mas yr$^{-1}$; $\mu_\delta$ -1.7 to -2.5 mas yr$^{-1}$; *Gmag* < 18.

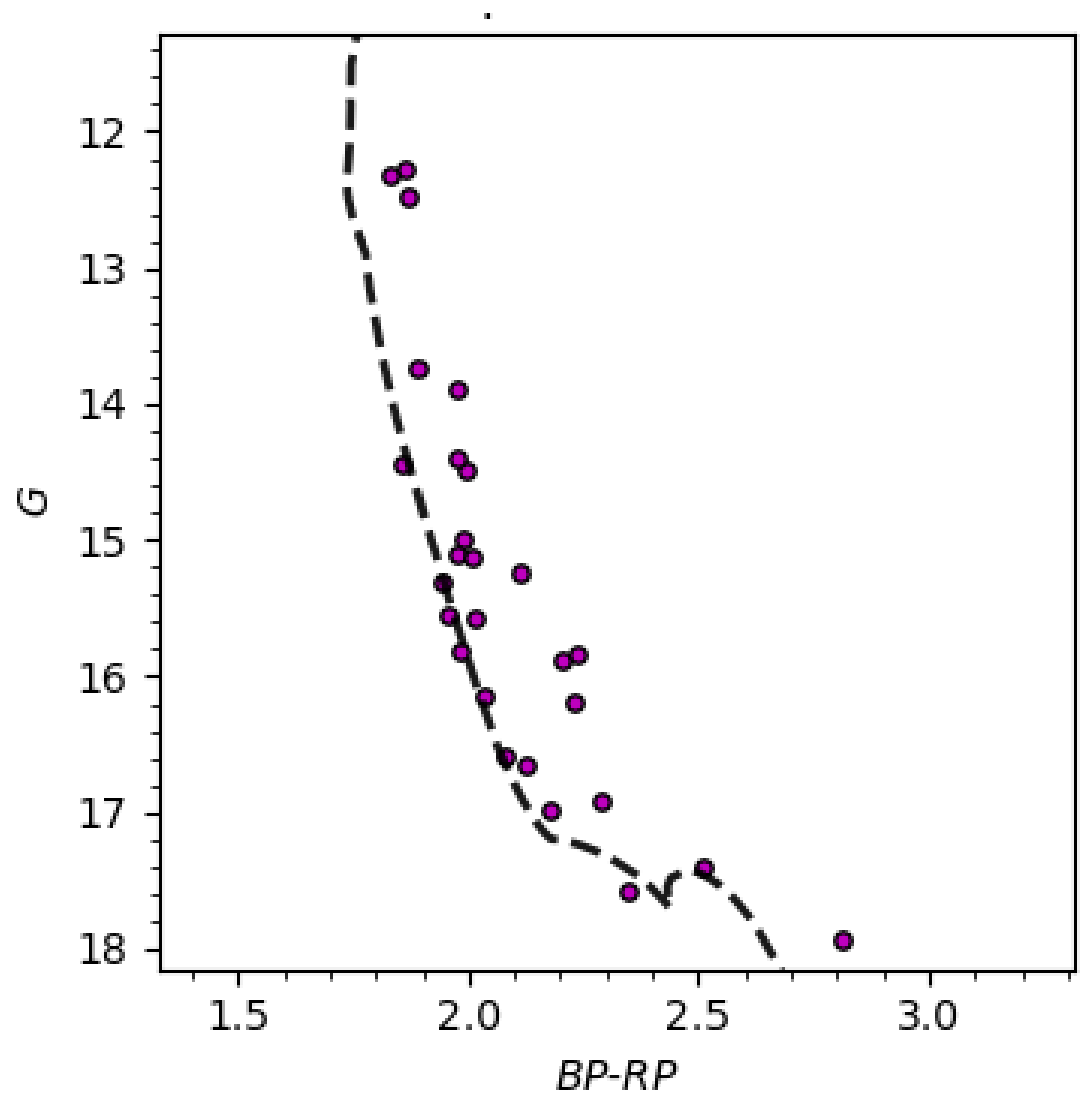

**Figure 2.** CMD of the most likely members of the new OC Casado 67 and the best fitting isochrone. Constraints are defined in Table 2.

NGC 6383 (Group D)

This candidate pair, formed by NGC 6383 and a newly identified sparse OC (Casado 68), appears associated with a giant molecular cloud that presumably is the nest where both were born not many Myr ago (Table 2).

NGC 6383 is a well-studied OC, which presents some substructure at its galactic north, suggesting partial disaggregation. Its reported RVs range from -1.2 km/s (Conrad et al. 2017) to 7.7 (Donor et al. 2020), comprising the mean value in Table 2. The cited range of RV fits the RV of two of the member stars of Casado 68 (Gaia EDR3 sources 4054489126483224704 and 4054444287024098048), which are 4.2 km/s and 4.6 km/s, respectively. The photometric distance of Casado 68 is 1.2 kpc (Fig. 3), compatible with the distance of NGC 6383 and with the mean parallaxes of both OCs. The extinction of the new OC is $A_V$ = 1.1 mag. The rest of their parameters in Table 2 are also well-matched. This agreement suggests the membership of both OCs to the same primordial group.

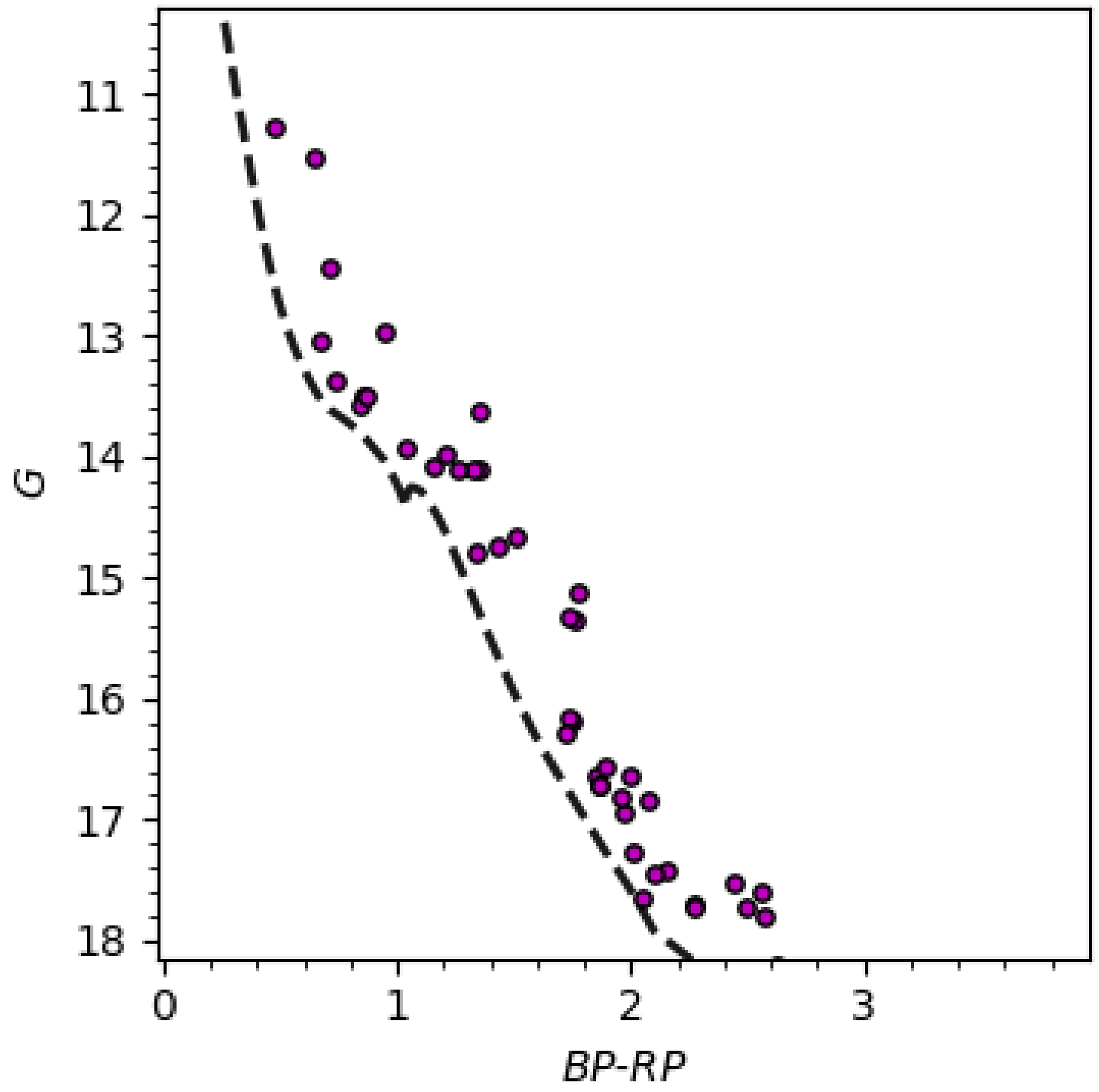

**Figure 3.** CMD of the most likely members of the new OC Casado 68 and the best fitting isochrone. Constraints are defined in Table 2.

FSR 0826

No OC linked to FSR 0826 has been found considering both its 3D position and 3D kinematics.

NGC 1893 (Group E)

NGC 1893, yet embedded in its parent molecular cloud, might be associated with the as yet unknown OC Casado 69. This new cluster is still embedded in the HII region IRAS 05197+3355, suggesting it is also young. The 6D astrometric parameters of both OC are well-matched, except the RV, which is unknown for the star members of Casado 69. However, a reported RV of -4.4 km/s for IRAS 05197+3355 (Wouterloot & Brand 1989) fits perfectly with the mean RV for NGC 1893 (Table 2), which supports the case for a genuine pair. In this regard, all reported RVs of NGC 1893 range from -2.2 km/s (Conrad et al. 2017) to -9.2 km/s (Kharchenko et al. 2013).

NGC 6193 (Group F)

NGC 6193 could be associated with the new OC Casado 70, apparently embedded in the same molecular cloud and partially covered by the dark nebula DOBASHI 6513. As shown in Table 2, all astrometric parameters match well, except for the RV, which is unknown for the likely stellar members of Casado 70. The photometric distance of the new OC is compatible with its mean parallax and with the distance to NGC 6193. The estimated extinction of Casado 70 is $A_V = 1.4$ mag. Both OCs appear to be young and have a common origin as their combined CMD fits quite well, considering the presence of associated nebulosity (Fig. 4). Moreover, the estimation of the age of Casado 70 by isochrone fitting confirms that preliminary assumption (Table 2).

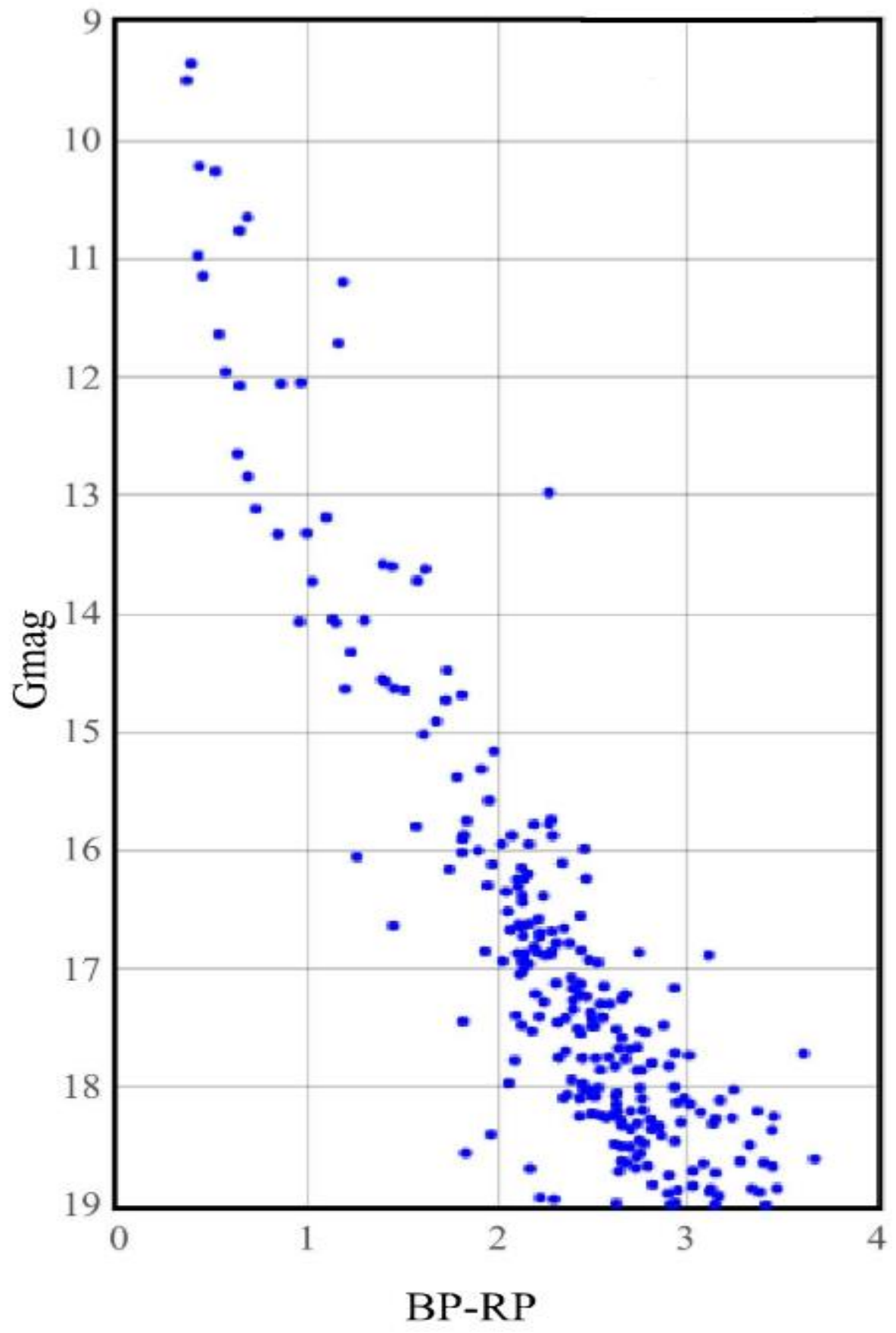

**Figure 4.** Combined CMD plot of the candidate primordial pair formed by NGC 6193 and Casado 70. Constraints: *plx* 0.77 to 0.95 mas; $\mu_\alpha$ 0.7 to 1.8 mas yr$^{-1}$; $\mu_\delta$ -4.0 to -4.8 mas yr$^{-1}$; *Gmag* < 19; *r* = 22 arcmin, centered at galactic coordinates 336.51, -1.39.

NGC 3572 (Group G)

This OC, adjacent to the parent molecular cloud [RC2004] G290.6+0.2-18.1, seems associated with the close OC Hogg 10. Moreover, some of the stars of Hogg 10 would belong to NGC 3572, according to data in Cantat-Gaudin et al. (2020). As summarized in Table 2, all *Gaia* derived parameters are consistent for both OCs. Reported distances to Hogg 10 range from 1.8 kpc (Loktin & Popova 2017) to 2.5 kpc (Glushkova et al. 1997). This range comprises the distance of NGC 3572 (Table 2). RVs of -7 km/s (Kharchenko et al. 2013) and 1 km/s (Dias et al. 2002) have been reported for Hogg 10, which are also compatible with that of NGC 3572 (Table 2). Last but not least, most reported ages are close to 6 Myr (Loktin & Matkin 1994; Dias et al. 2002; Ahumada & Lapasset 2007; Kharchenko et al. 2013; Loktin & Popova 2017; Hao et al. 2021), i.e., close to the age of NGC 3572. Thus, this pair fulfils all requirements of a primordial system.

Another candidate group member is the recently discovered cluster LP 1531 (Liu & Pang 2019), which we have confirmed as a genuine OC using *Gaia* EDR3. All relevant parameters in Table 2 are well-matched with the other Group G members, although we have not found any member star of LP 1531 with a measured RV.

FSR 0198 (Group H)

This OC appears to form a double system with Teutsch 8. All the relevant parameters are well-matched within the observational error (Table 2). However, no RV has been retrieved for Teutsch 8. This group could involve a third member: NGC 6871 (see below).

Pismis 27

Pismis 27 might be associated with the possible EC [KPS2012] MWSC 0759 (Kharchenko et al. 2013), but such a relationship is uncertain, as well as the definitive classification of [KPS2012] MWSC 0759.

UBC 568

No evidence of OC associated with UBC 568 has been found considering position, parallax and PMs.

NGC 7067

This NGC cluster has a second core of ca. ten stars at galactic coordinates 91.24, -1.72, which could either result from the fragmentation of a primordial OC or might be the outcome of a second OC born from the same parent molecular cloud. The apparent distance among both centres is ca. 5 arcmin, which corresponds to a projected distance of at least 10 pc, considering the mean parallax of NGC 7067 (0.15 mas; Cantat-Gaudin et al. 2020). This mean parallax is used since the reported distances are very assorted. No other associated candidates have been found within a radius of 100 pc, considering all the astrometric *Gaia* data.

NGC 2362 (Group I)

NGC 2362 seems to be associated with the EC Camargo 997, which is characterized here for the first time using *Gaia* EDR3. The mean parallax of Camargo 997 is less precise than usual due to the interference of nebulosity. However, all astrometric parameters are well-matched within the observational error (Table 2). Camargo 997 appears still partially embedded in the reflection nebulae Ced 96a and [RK68] 70, suggesting it is also a very young cluster. Both CMDs are also well-matched. A conceivable star member of this cluster (Gaia EDR3 5617723283645039616) has an RV of 30 km/s, although it is 13 arcmin apart of its apparent center. This RV accommodates within the reported RVs of NGC 2362: from 25 km/s (Vande Putte et al. 2010) to 36 km/s (Dias et al. 2002). The ensemble of data suggests that both clusters probably form a primordial system.

UBC 438 (Group J)

UBC 438, an elongated OC recently discovered by Cantat-Gaudin et al. (2020), seems associated with the new OC Casado 71. All known astrometric data are compatible (Table 2). Although some of the reported distances of UBC 438 may seem small considering the reported mean parallaxes, an updated estimation of 4.1 kpc (Dias et al. 2021) alleviates this minor discrepancy, possibly due to the high relative error of parallax. From the ensemble of data, the link between both OC looks plausible, but the actual existence of Group J requires confirmation.

NGC 6871 (Group $H_1$)

This Group could be related to Group H (see below). NGC 6871 is a large, rich OC with some substructure and at least two cores. Liu & Pang (2019) proposed NGC 6871 and Gulliver 17 as candidate members of an OC group. However, most *Gaia* astrometric parameters of both OCs, particularly the PMs, are discordant. On the other hand, NGC 6871 might be associated with Teutsch 8, even though distances and PMs are not identical (Table 2). For NGC 6871, the reported photometric distances range from 1.51 kpc (Glushkova et al. 1997) to 1.84 kpc (Cantat-Gaudin et al. 2018), while for Teutsch 8, literature distances span from 1.60 kpc (Dias et al. 2002) to 1.98 kpc (Cantat-Gaudin et al. 2020). Given their similar parallaxes, both OCs are likely to be at a compatible distance from the ensemble of data (~1.9 kpc). There is a consensus on the mean PMs derived from *Gaia* data, which seem coherent for both OCs. Quantitatively, ΔPM/plx (and ΔPM $d$) are ~0.9, which implies an acceptable difference in tangential velocities (< 5 km/s).

If NGC 6871 is related to Teutsch 8, it would also be associated to FSR 0198 (see Group H). The similar RVs of NGC 6871 and FSR 0198 from *Gaia* DR2 (Table 2) increase the likelihood of a triple Group. However, there is no literature consensus on the RV of NGC 6871. The ages of the three candidate members are again compatible with a unique (and recent) origin. Altogether, the case for a triple primordial group seems likely.

IC 1805 (Group K)

Astrometric *Gaia* data strongly suggest that IC 1805, a cluster embedded in its parent molecular cloud, is associated with Berkeley 65, even if a slight difference in $\mu_\delta$ is observed in Table 2. However, the photometric distances seem different at first sight. Nonetheless, most reported values of Berkeley 95 range from 1.9 kpc (Buckner & Froebrich 2013) to 2.28 kpc (Cantat-Gaudin et al. 2020), while for IC 1805, photometric distances span from 1.7 kpc (Kharchenko et al. 2013) to 2.34 kpc (Dias et al. 2002). Thus, considering the consistent *Gaia* parallaxes, both clusters could be at a distance of ~2.1 kpc. The ages of both OCs seem somewhat different (Table 2), but there is not a consensus on the age of Berkeley 65: This OC could be only 6 Myr old (Loktin & Matkin 1994; Ahumada & Lapasset 1995; Liu & Pang 2019).

The candidate cluster Camargo 755, partially embedded in nebula BRC 7, might also be a component of the same group. We have studied the field with *Gaia* EDR3 and confirmed the existence of a physical cluster, which astrometric mean parameters almost perfectly fit with those of Berkeley 65 (Table 2). The fact that Camargo 755 is an EC suggests that it is also very young. It is, therefore, a good candidate member for Group K. The EC MDF 10 (Bica et al. 2019), in giant HII region IC 1795, might be another member of the group, but this possibility requires confirmation.

VdBH 205 (Group L)

The parameters of VdBH 205 are like those of ESO 332-08 (Table 2), which suggest both clusters were born recently from the same molecular cloud where they are still embedded. The parallax of ESO 332-08 seems somewhat smaller, but a new value of 0.59 mas (Hao et al. 2021), based on *Gaia* EDR3, seems to solve this discrepancy and is roughly consistent with a global distance of ~1.7 kpc. Congruently, most reported values for VdBH 205 range from 1.54 kpc (Tarricq et al. 2021) to 2.16 kpc (Dias et al. 2002). The ages of both clusters are also similar. Thus, although no RV has been retrieved for ESO 332-08, our preliminary conclusion is that both objects are probable members of Group L. Unexpectedly, the OC UBC 323 (Cantat-Gaudin et al. 2020) seems to encompass some member stars of both OC.

The Infrared star cluster [DBS2003] 114 (Morales et al. 2013), still embedded in the HII region [CH87] 345.308+1.471, is also a firm candidate to be a member of the same group. The reported gas line velocity (-15 km/s) seems at odds with the RV of VdBH 205 in Table 2, but a compatible RV of -7 km/s has been frequently reported (Dias et al. 2002; Kharchenko et al. 2013; Conrad et al. 2017). The additional parameters retrieved using *Gaia* EDR3 are well-matched with their siblings (Table 2).

UBC 19

No linked OC has been found in a field 20 degree wide around this nearby cluster.

……..

In sum, we have found that 12 out of 20 studied young OCs (< 0.01 Gyr old) have at least one primordial companion. Three candidate Groups are dubious, and no companion has been found for five of the young clusters. Therefore, the total number of Groups ranges between 12 and 15 out of a total sample of 19 young OCs (SAI 25 has not been taken into account as it is not likely to be young). The resulting statistic suggests that the probability of the sample of young OCs of having linked siblings is 71±8 %. This ratio amounts to ca. six times the average fraction of linked clusters versus the total OC population in the Galaxy and the Magellanic Clouds, which is estimated at around 12% (Casado 2021 and references therein).

## 5. Grouping around old OCs

Conversely, 18 out of the 20 oldest OCs listed in Tarricq et al. (2021) turn out to be single OCs. All of them are believed to be older than 4 Gyr. These are, namely, Berkeley 17, NGC 188, NGC 6791, Collinder 261, NGC 1193, Berkeley 39, Trumpler 19, Berkeley 32 (an apparent link to Czernik 27 has been discarded), Berkeley 20, FSR 1521, Melotte 66, FSR 1407, NGC 2243, Haffner 5, Trumpler 5, ESO 092-05, King 11, and Berkeley 18. There are, however, two possible exceptions to this general rule:

Berkeley 36 (Group M)

Berkeley 36 might be associated with the new OC Casado 72. However, this affiliation requires confirmation since both parallaxes are only marginally compatible, even correcting for the general offset of Gaia DR2 parallaxes (Lindegren et al. 2018). The parallaxes, if correct, would imply that Berkeley 36 is ca. 400 pc farther away than Casado 72, and both OC would form merely an optical pair. The estimation of the photometric distance of the new OC, consistent with its mean parallax, endorses that assumption (Table 2). On the other hand, the difference in mean PMs between both objects (0.18 mas/yr) would correspond, at an assumed distance of 4 kpc, to a difference of 3.4 km/s in tangential velocity. Unfortunately, there is no known RV for the new OC. The CMDs of both OCs do not match, and accordingly, their ages are completely different (Table 2). Thus, Group M, if confirmed, would not be primordial. Nevertheless, from the ensemble of results, the physical existence of this double cluster appears very doubtful at present. The interstellar extinction of Casado 72 is $A_V$ = 1.6 mag.

Kronberger 81 (Group N)

Kronberger 81 may form a double system with Teutsch 17. The ensemble of the astrometric results seems to fit reasonably well. However, the photometric distances and ages are not alike. There is no consensus on the distance of Kronberger 81. Values span from 2.5 kpc (Loktin & Popova 2017) to 7.6 kpc (Buckner & Froebrich 2013). However, the mean value in Table 2 seems to be a reasonable compromise since it also agrees with the reported parallax. The photometric distance of Teutsch 17 (Table 2) appears too small for the corresponding parallax, but no other distance has been retrieved. If both OCs were roughly at the same distance from the Sun, say ~4 kpc, the distance between them would be >70 pc. The reported ages of Teutsch 17 (Table 2) are discordant, but both values are much lower than the age of Kronberger 81. Nonetheless, Kronberger 81 could also be 0.4 Gyr old (Dias et al. 2002; Kharchenko et al. 2013). Unfortunately, no RV has been retrieved for Teutsch 17. All in all, the physical link of this candidate pair is uncertain.

Considering all these cases, the probability of any old cluster being part of a binary system appears to be <10% and, most likely, is close to 0.

The theoretical models indicate that binary cluster lifetimes range from a few Myr to ca. 0.04 Gyr (e.g., Bhatia 1990). Congruently, Grasha et al. (2015) also found that in the galaxy NGC 628, the clustering of star clusters decreases very rapidly with cluster age for OCs older than 0.04 Gyr. Therefore, we consider a simplified bimodal model where young OC (< 0.04 Gyr) have the above-derived probability of 71±8 % of forming part of a group and older OCs have a very low likelihood (~ 0) of being part of a group. When we apply this model to the total sample of clusters from Tarricq et al. (2021) that have an estimated age (1315 clusters), counting the young clusters (229) separately, the result is an overall probability of (229/1315) 71±8 % = 12.4±2 % that any cluster is linked to another cluster. This value is remarkably close to the above-mentioned average fraction of linked clusters in the Galaxy (Casado 2021). If we apply the same model to another updated catalogue (Hao et al. 2021), the result obtained is similar (16±2 %). In both cases, the quoted error estimates (2 %) are scaled from the original probability (71±8 %) and rounded up. So, a reasonable overall fraction of associated OCs is obtained by assuming that OCs younger than 0.04 Gyr are still most likely associated with their primordial relatives while older OCs are most likely isolated since their siblings have been separated or disintegrated by tidal forces in the Galaxy and close encounters with giant molecular clouds (Bhatia 1990). Since the obtained figures are similar for the Galaxy and the Magellanic Clouds, we suggest that this mechanism could be general in disc galaxies where OCs are formed. Our findings agree, at least qualitatively, with those of Bica et al. (2003) for ECs in the Galaxy (mentioned in the Introduction) and with the pioneering work of Larsen (2004), who studied young OCs in nearby spiral galaxies. Larsen found that many of the youngest objects are in very crowded regions, and approximately 1/3–1/2 of them are double or multiple sources.

**6. The double cluster in Perseus and other reported binary cluster candidates**

As stated in the introduction, the classical double cluster formed by h and $\chi$ Persei (NGC 869 and NGC 884) was the only confirmed physical pair known in our Galaxy until recently. Let's review their characteristics to see if they would also be considered an actual binary cluster according to our selection criteria. From the angular distance among both OC centres (27 arcmin), and assuming that they are at the same heliocentric distance (2.2 kpc; Cantat-Gaudin et al. 2020), an estimated distance among them of ca. 20 pc can be inferred, well within the 100 pc limit that we use as a selection criterion. The median reported RV for NGC 869 (h Persei) is -42.8 km/s (Dias et al. 2002), and for NGC 884 ($\chi$ Persei) is -43 km/s (Dias et al. 2002; Loktin & Popova 2017; Soubiran et al. 2018), which are practically coincident. The *Gaia* DR2 mean parallaxes are 0.399 mas and 0.398 mas, respectively (Cantat-Gaudin et al. 2020). Thus, $\Delta$plx/plx is less than 1%. Using PMs from Cantat-Gaudin et al. (2020), $\Delta$PM/plx (and $\Delta$PM $d$) is ≤ 0.2 yr$^{-1}$. Consequently, the resulting orbital elements of both OCs are companionable within a 1$\sigma$ deviation interval (Tarricq et al. 2021). Therefore, it is clear that the binary system formed by NGC 869 and NGC 884 fulfills all our adopted criteria for a well-behaved binary system, but… what about the age?

According to the *Primordial group* hypothesis, this physical pair should be of primordial origin, i.e., formed by young clusters of comparable age. For NGC 869, the extensive literature quotes ages from 3 Myr (Dambis 1999) to 19 Myr (Kharchenko et al. 2013). However, there is a broad consensus on the mean age of 12 Myr (e.g., Loktin & Matkin 1994; Dias et al. 2002; Ahumada & Lapasset 2007; Vande Putte et al. 2010; Loktin & Popova 2017; Rain et al. 2021). NGC 884 has practically the same span on reported ages: from 3 Myr (Ahumada & Lapasset 1995) to 18 Myr (Cantat-Gaudin et al. 2020), but many reports converge around 13 Myr (Dias et al. 2002; Wu et al. 2009; Vande Putte et al. 2010; Gozha et al. 2012; Hayes & Friel 2014; Loktin & Popova 2017; Rain et al. 2021). So, both OCs are indeed young and have almost the same age within the observational error margins, as expected from the *Primordial group* hypothesis.

Some other candidate binary clusters having at least one member older than 0.1 Gyr have been reported in the literature. Let's reexamine a few of them:

NGC 2383/NGC 2384

This pair was proposed as a probable binary cluster by Subramaniam et al. (1995). Kopchev et al. (2008) concluded that these two OCs were not born in the same molecular cloud, given their significant age difference.

Whatever the case, both NGC clusters do not appear to be physically associated, given their disparate characteristics (except their close position in the sky). As detailed in Table 2, *Gaia* parallaxes and photometric distances agree that NGC 2383 is significantly farther away than NGC 2384. The ranges of reported RVs for both OCs do not overlap, confirming that they are not associated: RVs of NGC 2363 span from 55 km/s (Huang & Gies 2006) to 72 km/s (Tarricq et al. 2021; Dias et al. 2021), whilst RVs of NGC 2364 range from 46 km/s (e.g., Conrad et al. 2017) to 53 km/s (Dias et al. 2002; Garmany et al. 2015). Therefore, the diverse ages of this optical pair are not surprising, despite a minimum age of 15 Myr has also been reported for NGC 2383 (Ahumada & Lapasset 2007).

NGC 2384 is quite elongated (Bica et al. 2019) and has an extended halo (at least 40 arcmin) beyond its reported radius towards the north. By the way, this halo (encompassing NGC 2384) has been identified as the new OC UBC 224 (Cantat-Gaudin et al. 2020). Most reported ages range from 0.006 Gyr (Dambis 1999) to 0.021 Myr (Dias et al. 2021), i.e., it is a young OC by any standards. Unexpectedly, a new OC of compatible parameters has been found while studying the star field around NGC 2384. This cluster, Casado 73, seems somewhat older (0.08 Gyr), but it is also a young OC. The mean parallaxes of NGC 2384 and Casado 73 are somewhat discrepant (Table 2), but *Gaia* parallaxes of 0.33 mas (Dias et al. 2021) and 0.35 mas (Hao et al. 2021) for NGC 2384 alleviate such discrepancy. These parallaxes are compatible with the photometric distance of Casado 73 (3.1 kpc), which in turn is harmonious with its *Gaia* EDR3 parallax. Moreover, that distance is within the range of most reported photometric distances for NGC 2384: From 2.1 kpc (e.g., Loktin & Popova 2017) to 3.2 kpc (e.g., Glushkova et al. 1997). Furthermore, their CMDs show a coincident main sequence suggesting similar distance, age, and metallicity (Fig. 5). The estimated extinction of this new OC is $A_V = 0.51$ mag. The ensemble of data indicates that this is a plausible primordial pair (Group O), even though no RVs have been found for the member stars of Casado 73. Anyhow, this primordial pair candidate requires confirmation.

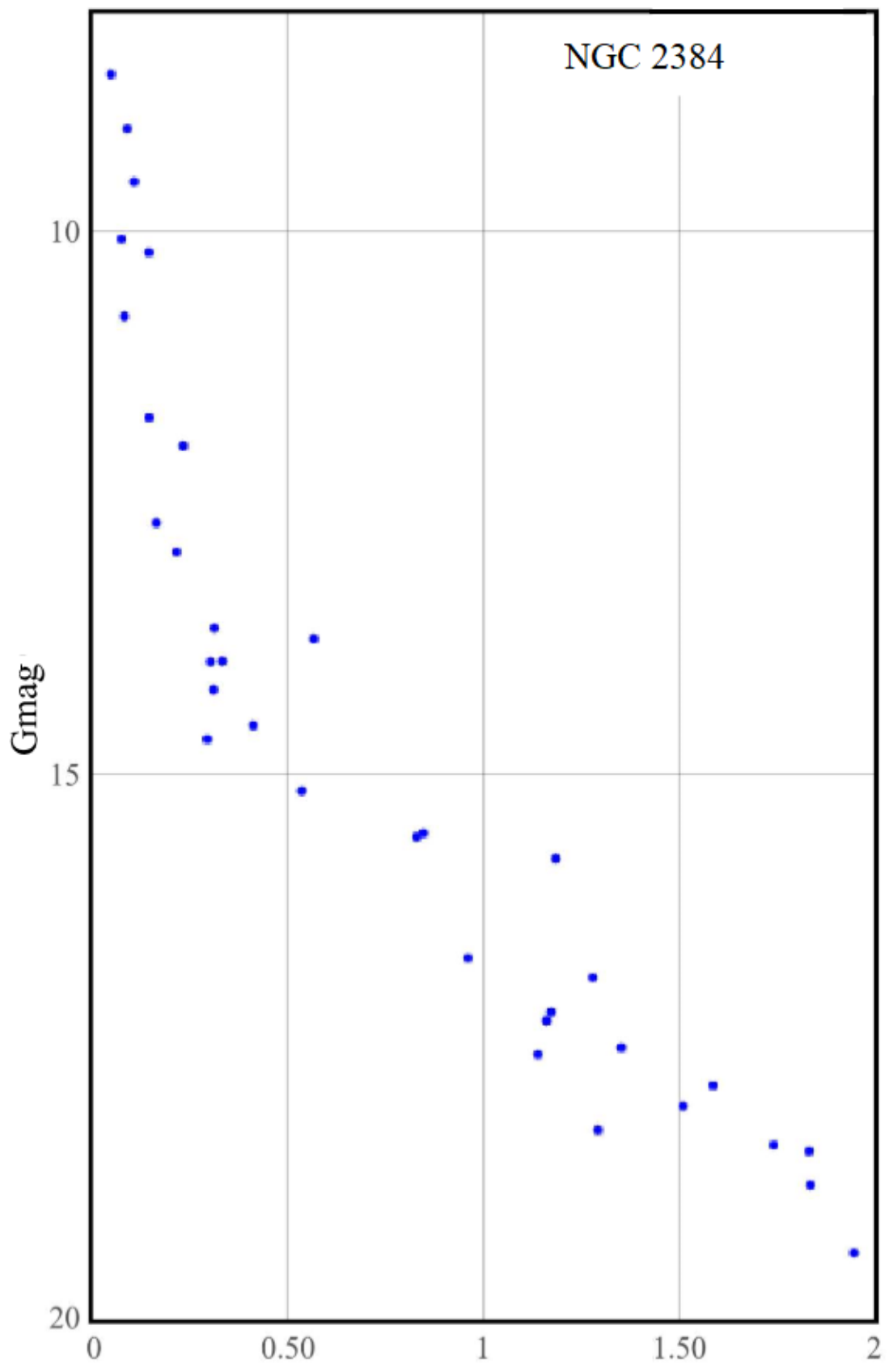

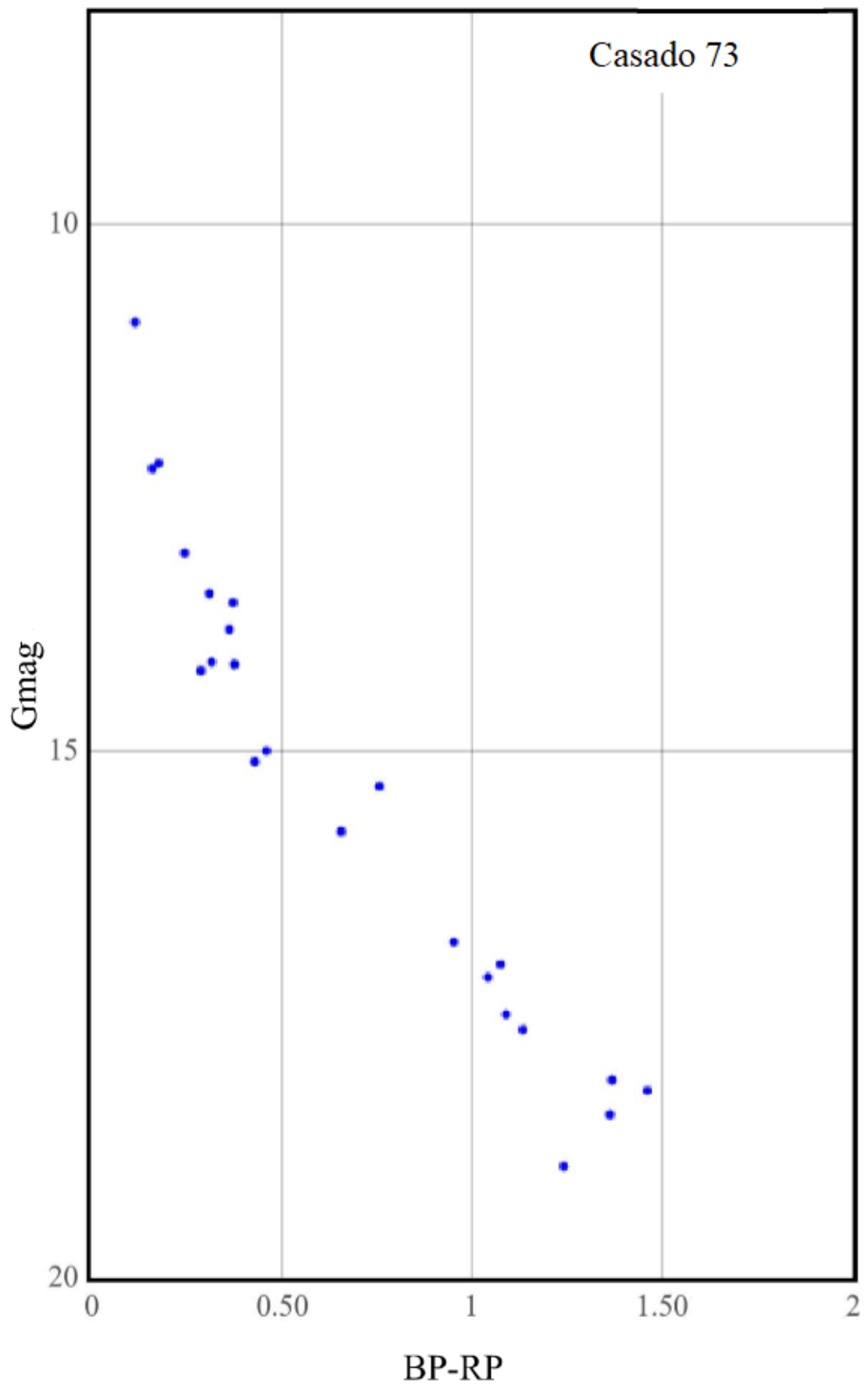

**Figure 5.** Compared CMD plots of the candidate primordial pair formed by NGC 2384 and Casado 73. Constraints for each OC are given in Table 2.

FSR 1767/Ruprecht 127

De la Fuente Marcos et al. (2014) proposed this pair of star clusters to be the result of the capture of Ruprecht 127 by the candidate globular cluster FSR 1767 (2MASS-GC04) (Bonatto et al. 2007). Despite FSR 1767 has been described as an OC using *Gaia* EDR3 (Hao et al. 2021), no other Gaia studies on it have been retrieved (see comment on pair #16b), and we have not found any trace of such a cluster in *Gaia* data. Previous literature only reported a single intermediate age (0.473 Gyr), although frequently recited (e.g., Dias et al. 2002; Kharchenko et al. 2013; Dib et al. 2018; Hao et al. 2021), and two discordant photometric distances: 1948 pc (e.g., Dias et al. 2002; Kharchenko et al. 2013; Dib et al. 2018) and 3.6 kpc (Buckner & Froebrich 2013). The number of reported member stars also varies greatly: from 17 (e.g., Dias et al. 2002) to 984 (Froebrich et al. 2007). All these conflicting results make very doubtful the physical existence of such OC, in which case no binary cluster containing FSR 1767 would exist.

…………..

Last but not least, all the likely binary clusters lately proposed by Soubiran et al. (2018, 2019) using *Gaia* data, namely ASCC 16/ASCC21, Collinder 140/NGC 2451B, IC 2602/Platais 8, RSG7/RSG 8, and Collinder 394/NGC 6716, appear to be young (≤ 0.1 Gyr) and have ages compatible with a common origin. Their group of five members containing ASCC 16, ASCC 19, ASCC 21, Gulliver 6, and NGC 2232 also follow the same rules and, thus, appear to form a primordial group.

## 7. Concluding remarks

In this work, we formulate and test the *Primordial group* hypothesis. It states that OCs are born in primordial groups that disperse through the galactic disc in a relatively short time (≤ 0.1 Gyr). We test that hypothesis through manual mining of *Gaia* EDR3 and careful revision of the extensive literature on OCs.

The revision of candidate pairs in DFM with at least one of their members older than 0.1 Gyr allows us to conclude that practically all of them can be discarded as actual binary clusters. Some of the pairs are optical pairs, others are hyperbolic encounters, and a few pairs may be primordial pairs with flawed ages. A significant number of the clusters studied in this sample are most likely false OCs: Loden 1171, ASCC 34, Loden 565, VdBH 91, ASCC 4, NGC 1746, Basel 5, Loden 165, ESO 128-16 and ESO 132-14.

We revisited the twenty youngest OCs (< 0.01 Gyr) listed by Tarricq et al. (2021), looking for associated clusters closer than 100 pc and sharing PMs and RVs. The resulting statistics suggest that the probability that young OCs have related siblings is 71±8 %. On the other hand, the probability that older OCs (> 4 Gyr) from the same catalogue are not alone seems very low, if not zero. A reasonable overall fraction of associated OCs (12-16%) is obtained from a simplified bimodal model, which assumes that OCs younger than 0.04 Gyr are still most likely associated with their primordial relatives, while older OCs are most likely isolated. However, these proportions are only approximate due to the incompleteness of the sample.

Seven new OCs have been identified during this research (namely, Casado 67-73). This unexpected result reveals that the search for associated clusters around very young stellar clusters is an effective method for discovering new OCs, as is the search for new OCs around a given grouping (Casado 2021).

The classical double cluster in Perseus fulfills all our selection criteria for a binary system. Both members are indeed young (< 0.02 Gyr) and have practically the same age, as expected from the tested hypothesis. Some other reported binary cluster candidates with putative members greater than 0.1 Gyr have been reasonably discarded. On the other hand, the likely OC groups from Soubiran et al. (2018, 2019) are young and compatible with the *Primordial group* hypothesis.

Three of the revised OCs (UBC 224, UBC 323 and UBC 344; Castro-Ginard et al. 2020) have been found to encompass a significant number of member stars of other well-known OCs.

The present results indicate that the vast majority of real double/multiple OCs in the Galaxy, if not all, are of primordial origin and are not stable for a long time, in line with similar conclusions obtained from the study of the Magellanic Clouds (Hatzidimitriou & Bhatia 1990; Dieball et al. 2002). Thus, the pairs of OCs in these groups are generally not true binary systems since they are not gravitationally bound. The *Primordial group* hypothesis has successfully passed these four tests and, therefore, deserves further scrutiny as a feasible working model.

**Acknowledgments:** The author is indebted to David Jou for his longstanding support and encouraging discussions. Thanks to Friedrich Anders and Alfred Castro Ginard for providing photometric distances and ages of some of the new OCs found in this study. This work has made use of data from the European Space Agency (ESA) mission *Gaia* (https://www.cosmos.esa.int/gaia), processed by the *Gaia* Data Processing and Analysis Consortium (DPAC, https://www.cosmos.esa.int/web/gaia/dpac/consortium). Funding for the DPAC has been provided by national institutions, in particular the institutions participating in the *Gaia* Multilateral Agreement. This research made extensive use of the SIMBAD database, and the VizieR catalogue access tool, operated at the CDS, Strasbourg, France (DOI: 10.26093/cds/vizier), and of NASA Astrophysics Data System Bibliographic Services.

**Data availability statement:** The data associated with this manuscript are available in the public data repositories mentioned in the Acknowledgements section and in the referenced studies.